\documentclass[structabstract]{aa} 
\usepackage{graphicx}
\usepackage{lscape}
\usepackage{natbib}
\usepackage{txfonts}
\usepackage{longtable}

\newcommand{\kms}{km~s$^{-1}$}
\newcommand{\Mo}{M$_{\odot}$}
\newcommand{\Teff}{T_{\rm eff}}
\newcommand{\FeH}{[Fe/H]}
\newcommand{\sFe}{[s/Fe]}

\newcommand{\bacchus}{{\footnotesize BACCHUS}}
\newcommand{\marcs}{{\footnotesize MARCS}}

\begin{document}
\title {To Ba or not to Ba: Enrichment in s-process elements in binary systems with WD companions of various masses}

\author{T. Merle\inst{1}
   \and A. Jorissen\inst{1}
   \and S. Van Eck \inst{1}
   \and T. Masseron\inst{1,2}
   \and H. Van Winckel\inst{3}
}
\institute{Institut d'Astronomie et d'Astrophysique, Universit\'e Libre de Bruxelles, CP 226, Boulevard du Triomphe, 1050 Brussels, Belgium \\
           \email{tmerle@ulb.ac.be}
\and
Institute of Astronomy, University of Cambridge,  Madingley Road, Cambridge CB3 0HA, UK
\and
Instituut voor Sterrenkunde, Katholieke Universiteit Leuven, Celestijnenlaan 200D, 3200 Heverlee, Belgium
}

\date{Received ...; accepted ...}

\abstract
{The enrichment in s-process elements of barium stars is known to be due to pollution by mass transfer from a companion formerly on the thermally-pulsing asymptotic giant branch (AGB), now a carbon-oxygen white-dwarf (WD).}
{We are investigating the relationship between the s-process enrichment in the barium star and the mass of  its WD companion. It is expected that helium WDs, which have masses smaller than about 0.5~\Mo\ and never reached the AGB phase,
should not pollute with s-process elements their giant companion, which should thus never turn into a barium star.}
{Spectra with a resolution of $R\sim86\;000$ were obtained with the HERMES spectrograph on the 1.2-m Mercator telescope for a sample of 11 binary systems involving WD companions of various masses. We use standard 1D LTE \marcs\ model atmospheres coupled with the Turbospectrum radiative-transfer code embedded in the \bacchus\ pipeline to derive the atmospheric parameters using equivalent widths of \ion{Fe}{i} and \ion{Fe}{ii} lines. Least-square minimization between the observed and synthetic line shape is used to derive the detailed chemical abundances of CNO and s-process elements.}
{The abundances of s-process elements for the entire sample of 11 binary stars were derived homogeneously. The sample encompasses all levels of overabundances: from solar \sFe$=0$ to $1.5$~dex in the 2 binary systems with S-star primaries (for which dedicated MARCS model atmospheres were used).  The primary components of binary systems with a WD more massive than 0.5~M$_\odot$ are enriched in s-process elements. We also found a 
 trend of increasing \sFe\ with [C/Fe] or [(C+N)/Fe].}
{Our results conform to the expectation that binary systems with WD companions less massive than 0.5~\Mo\ do not host barium stars.}

\keywords{Stars: abundances -- white dwarfs -- Stars: late-type -- binaries: spectroscopic}

\titlerunning{To Ba or not to Ba?}

\maketitle
\section{Introduction}

The puzzle of the barium stars, a class of K giants with overabundances of heavy elements produced by the s-process of nucleosynthesis \citep{Bidelman-Keenan-51}, has been solved by \citet{McClure-80} who unravelled the binary nature of these stars, with the companion endorsing the responsibility of the 
heavy-element synthesis.  This solution of course requires the companion to be a carbon-oxygen white dwarf (WD), which synthesized the heavy elements 
while on the thermally-pulsing asymptotic giant branch (TP-AGB). An indirect argument supporting that hypothesis has been provided by 
\citet{McClure-Woodsworth-90} and \citet{Jorissen1998} in the form of the mass-function distribution, consistent with the companion masses being peaked at 0.6~\Mo, as expected for carbon-oxygen WDs. It is only quite recently that a clear {\it direct} argument was presented by \citet{Gray2011}, in the form of the detection by GALEX of UV excess flux  from those systems, attributable to a warm WD. Earlier studies by \citet{Schindler-82}, \citet{Dominy-Lambert-1983}, \citet{Bohm-Vitense-84,BohmVitense00}, \citet{Jorissen-96},  and \citet{Frankowski-2006}
found similar evidence, although somewhat less convincingly or for a smaller set.

With the advent of sensitive far-UV and X-ray satellites, like EUVE, XMM-Newton and ROSAT, many warm WDs were discovered from their X-ray or UV flux
 \citep[see][and references therein]{Bilikova2010}. Some among these are members of binary systems involving a late-type primary star \citep[e.g.,][]{Landsman1993,Vennes1995,Vennes1997,Vennes1998,Jeffries-Smalley-96,Jorissen-96,Christian1996,Burleigh1997,Burleigh1998,Hoard2007,Bilikova2010}, 
with the latter thus being barium-star candidates.  
Of course, for this to happen, the progenitor of the current WD had to be a TP-AGB star with an atmosphere enriched in s-process elements. This in turn requires that the s-process nucleosynthesis 
operated in the H- and He-burning  intershell region  {\it and} that the s-process material was dredged up to the surface. The currently available sample of WDs in binary systems offers an interesting way to test the above paradigm, especially when the mass of the WD is known. By checking whether or not the {\it current companion} of the WD is enriched in s-process  elements, we may trace the operation of s-process and dredge-up in the AGB progenitor of the current WD. Since its mass is equivalent to the AGB core mass at the end of its evolution, we may thus set constraints on models of AGB evolution by investigating whether there is a threshold on the WD mass for its companion to become a barium star. It is expected for instance that He WDs, which are less massive than CO WDs, had progenitors which did not ascend the AGB. Their companion should thus never be barium stars. This was recently proven to be true for the IP~Eri system, whose K primary star, companion of a He WD, is not a barium star  \citep{2014A&A...567A..30M}. The present paper addresses this issue on a larger sample of binaries involving WDs covering a broad range of masses. 

The paper is organized as follows. Sect.~\ref{Sect:sample}   describes the stellar sample and Sect.~\ref{Sect:observations} the observations. Sect.~\ref{Sect:atmosphere} presents the method used for deriving the atmospheric parameters, and Sect.~\ref{Sect:abundances} the resulting abundances, separately for CNO and s-process elements. Sects.~\ref{Sect:discussion} and \ref{Sect:conclusion} 
contain the discussion about these results, and our conclusions, respectively.

\begin{table*}
 \caption{The sample of binary systems with a WD companion and known orbital elements ($P$ and $e$ are the orbital period and eccentricity, respectively, and $f(M_1,M_2)$ is the mass function), ordered by increasing WD mass.}
 \small
 \center
\begin{tabular}{lllllll}

Names  & Spectral type & $M_{\rm WD}$ (\Mo)& $P$ (d)& $e$ & $f(M_1,M_2)$ (\Mo) & Ref.\\
\hline

\medskip\\
EUVE J0459-102          &  G4V + He WD  & 0.3                & $903\pm5$            & $0.30\pm0.06$      & $0.009\pm0.003$            & 2, 3\\
HR 1608  (63 Eri)       &               &                    &                      &                    &                            &
\medskip\\
HD 185510 (V1379 Aql)   & K0III + sdB   & $0.30 - 0.36$      & $20.66187\pm0.00058$ & $0.094\pm0.011$    &  $0.0042\pm0.0002$         & 1
\medskip\\
EUVE J0254-053          & K0 IV + He WD & 0.4                & $1071.0\pm1.8$       & $0.25\pm0.01$      & $0.0036\pm 0.0001$         & 4, 5\\
HD 18131 (IP Eri)       &               &                    &                      &                    &                            &
\medskip\\
EUVE J0515+326          & F3V           & $0.53 - 0.69$      & $2.99322\pm0.00005$  & 0                  & $0.0079\pm0.0002$          & 3\\
HD 33959C (14 Aur C)
\medskip\\  
EUVE J1732+74.2         & K0III         & 0.55               & $903.8\pm0.4$        & $0.07\pm0.03$      & $0.0035\pm0.0003$     & 1, 17\\
HD 160538 (DR Dra)
\medskip\\
HD 35155  (V1261 Ori)   & S4,1          & $0.55\pm0.05$            & $640.7\pm2.7$        & $0.07\pm0.03$      & $0.032\pm0.003$     & 6, 7\\
                        &               &                    &                      &                    &                            & 
\medskip\\  
1RXS J151824.6+203423   & G8III Ba0.3   & $0.59 - 0.79$      & $506.45\pm0.18$      &  $0.3353\pm0.0056$ & $0.0112\pm0.0003$ & 6, 8 \\
HR 5692                 &               &                    &                      &                    &                            &  
\medskip\\
HD 121447 (IT Vir)      & K7III Ba SC2  & $0.5 - 0.7$        & $185.66\pm0.07$      & $0.0$              & $0.025\pm0.001$            & 16 
\medskip\\
HD 202109 ($\zeta$ Cyg) & G8III Ba0.6   & 0.7 - 1.1              & $6489\pm31$        & $0.22\pm0.03$      & $0.023\pm0.003$            & 9, 10, 14\\
                        &               &                    &                      &                    &                            & 
\medskip\\
RX J2307.1+2528         & K0Iab Ba      & 0.75 - 1.15          & $111.14\pm0.01$      & 0                  & $(3.7\pm0.3)\;10^{-5}$     & 11, 12, 13\\
HR 8796 (56 Peg)        &               &                    &                      &                    &                            &
\medskip\\
$\zeta$ Cap             & G4Ib Ba       & 1                  & $2378\pm55$        & $0.28\pm0.07$      & $0.004\pm0.001$            & 14, 15\\
HD 204075               &               &                    &                      &                    &                            & \medskip\\

\hline
\end{tabular}
\tablebib{
	(1) \citet{Fekel1993};
	(2) \citet{Landsman1993};
	(3) \citet{Vennes1998}
	(4) \citet{Burleigh1997};
	(5) \citet{2014A&A...567A..30M};
	(6) \citet{Jorissen-96};
	(7) \citet{Jorissen1992};
    (8) \citet{Stefanik2011};
    (9) \citet{Dominy-Lambert-1983};
    (10) \citet{Griffin1992};
    (11) \citet{Schindler-82};
    (12) \citet{Griffin2006};
    (13) \citet{Frankowski-2006};
    (14) \citet{Bohm-Vitense1980};
    (15) \citet{Jorissen1998};
    (16) \citet{Jorissen1995};
    (17) \citet{1985AJ.....90..812F}.
}
 \normalsize
  \label{tab:comp}
\end{table*}

\section{The sample}
\label{Sect:sample}

The binary systems selected for the present study have WD components orbiting  a cool star (with $\Teff<~7000$~K). The main selection criterion was the knowledge of the WD mass, which ranges from 0.3 to 1.0~\Mo. Our sample is listed in Table~\ref{tab:comp}. Incidentally, the sample covers a broad range of orbital periods and eccentricities.  The WD masses are taken from the references listed in Table~\ref{tab:comp}. 

Here follow  remarks on some targets of our sample.

IP Eri (HD~18131) consists of a K0~IV star and a He WD companion. Its evolutionary history has been studied by \citet{2014A&A...565A..57S}, whereas an abundance study was performed by \citet{2014A&A...567A..30M}. 

14~Aur~C (HD~33959) is member of a quadruple system with component~A being a $\delta$~Scuti star. 14~Aur~C is located 14.6~arcsec from A, at a distance of $104\pm32$~pc from the Sun \citep{2007A&A...474..653V}. Spectral types F2V+DA \citep{Vennes1998} or F5V \citep{Gray2003} have been proposed. It appears to be a fast rotator. We derived for this star $v\sin{i}=16$~\kms. No atmospheric parameters are available yet in the literature for that star. The orbital period of only 
3~d makes the dwarf star 14 Aur~C an outlier among the period distribution of giant barium stars, but not among that of dwarf carbon stars (should 14 Aur~C turn to be so), where at least one star with a similar period is known \citep[HE~0024-2523 has a period of 3.4~d;][]{2003AJ....125..875L}.

V1261 Ori (HD~35155) is a symbiotic S star \citep{1991ApJ...383..842A,Jorissen-96}. 

IT Vir (HD~121447) is a border case between barium and S stars, since it has been classified as SC2 \citep{Ake79}, K4Ba \citep{Abia98}, K7Ba and S0. 

$\zeta$~Cap (HD~204075) is a prototype barium star. 

\begin{figure}[h]
 \hskip-0.4cm \includegraphics[angle=-90,width=1.1\linewidth]{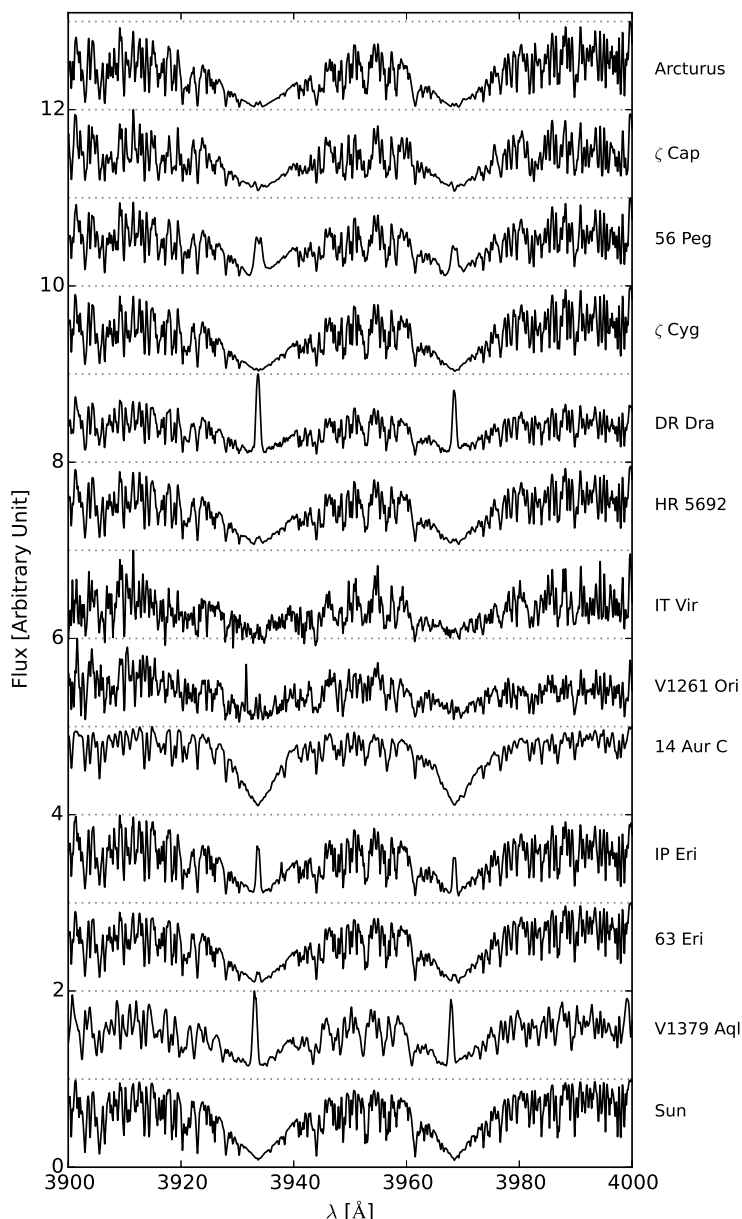}
 \caption{ \label{fig:spec_hk}
The Ca II H and K lines for the sample stars. For comparison, the spectra of the Sun and Arcturus are plotted at the bottom and at the top of the figure, respectively. Evidence of activity is present in 63~Eri, IP~Eri, V1379~Aql, DR~Dra and 56~Peg.}
\end{figure}

DR~Dra (HD~160538) and V1379~Aql (HD~185510) have strong emission cores in the \ion{Ca}{II} H and K lines (Fig.~\ref{fig:spec_hk}), indicative of chromospheric activity related to rapid rotation, as confirmed by our analysis ($v\sin{i}=8.5$~\kms\ and 18~\kms, respectively; see Fig.~\ref{fig:spec_baii}) in agreement with the results of \citet{Fekel1993}. Emission is also visible in the IR \ion{Ca}{II} triplet.

\begin{figure}[h]
 \label{fig:ap}
  \includegraphics[width=\linewidth]{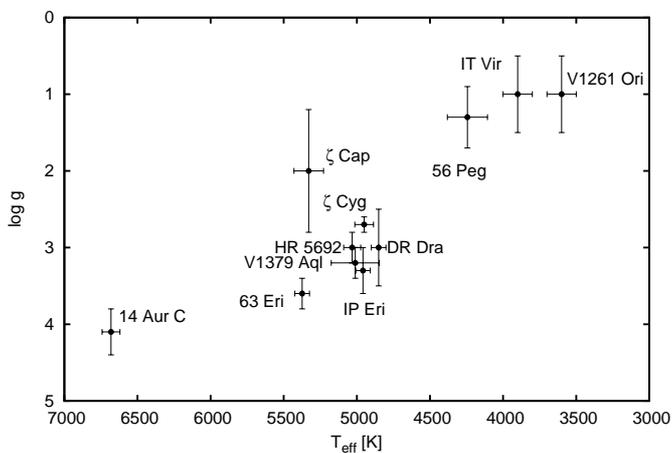}
 \caption{Surface gravity vs effective temperature for the sample stars.}
\end{figure}

\section{Observations}
\label{Sect:observations}
High-resolution spectra ($R\sim86\;000$) were obtained using the HERMES spectrograph \citep{Raskin2011} operating at the 
1.2~m telescope Mercator at La Palma (Canary Islands, Spain). The HERMES spectrograph is a fiber-fed echelle spectrograph which provides a full spectral coverage from 377 to 900~nm. All the spectra were reduced using the HERMES pipeline software and have a typical signal-to-noise ratio of about 200. The observations were carried out between June 2009 and September 2014.

\section{Atmospheric parameters}
\label{Sect:atmosphere}

First guesses for the atmospheric parameters were determined from previous works (as listed in Sect.~\ref{Sect:sample}), from photometry or spectral classification. 
We assume the metallicity to be solar to start with [the solar abundance set from \citet{Grevesse07} has been used throughout this paper, with in particular 
$A_\odot(\mathrm{Fe}) = 7.45$\footnote{defined as $A(\text{Fe}) = \log{[n(\text{Fe})/n(\text{H})]} + 12$, where $n(\text{Fe})$ and $n(\text{H})$ are the number densities of Fe and H.}]. 

The atmospheric parameters were then determined iteratively using the pipeline \bacchus\ developed by one of the author (TMa; see also \citealt{Jofre-etal-13}).
This pipeline is based on the 1D LTE spectrum-synthesis code Turbospectrum \citep{Alvarez1998,Plez2012} and the method to derive the atmospheric parameters is based on the fit of the equivalent widths of \ion{Fe}{i} and \ion{Fe}{ii} lines. The code derives automatically the effective temperature $\Teff$, the logarithmic surface gravity $\log{g}$, the global metallicity [Fe/H] and the microturbulent velocity $\xi$ by removing any trend in the \FeH\ vs. $\chi$ and \FeH\ vs. $W/\lambda$  relations (where $\chi$ is the energy of the lower level of a line, $W$ the measured equivalent width and $\lambda$ the wavelength of the considered line), and by forcing lines of  \ion{Fe}{I} and \ion{Fe}{II} to yield the same abundance. 
We used \marcs\ model atmospheres \citep{Gustafsson2008} and a line list constructed from the VALD3 database \citep{Kupka2011}, complemented for Fe and V  lines by  data from \citet{DenHartog2014a,DenHartog2014b} and \citet{Ruffoni2014}.

The synthetic spectra were convolved with a Gaussian function with a $\sigma$ varying from  6 to 12~\kms\ to match the stellar macroturbulence and/or HERMES resolution. 
Only reduced equivalent widths ($W/\lambda$) of \ion{Fe}{i} and \ion{Fe}{ii} lines lower than 0.025 are retained in the analysis.
For the fast rotators 14~Aur~C ($V \sin i = 16$~\kms) and V1379 Aql ($V \sin i = 18$~\kms), we convolved the synthetic spectra with a rotational profile.

\citet{2014A&A...567A..30M} performed an abundance analysis of IP~Eri, and obtained the following atmospheric parameters: $\Teff= 4960\pm100$~K, $\log g = 3.3\pm0.3$, and [Fe/H]$= +0.09\pm0.08$, which will be used here as well.

The atmospheric parameters of the prototypical Ba star $\zeta$~Cap ($\Teff=5397\pm82$~K, $\log{g}=1.48\pm0.16$, and $[\mathrm{Fe/H}]=-0.14\pm0.10$) were derived by  \citet{Prugniel2011}.
They were re-derived in the present study,  and the two parameter sets are close to each other.

For the two S stars (IT~Vir and V1261~Ori), the atmospheric parameters cannot be derived with \bacchus\ due to the presence of strong molecular lines which blend many Fe lines and prevent the use of their equivalent widths. Instead,  for V1261~Ori, we adopted the parameters derived by \citet{Neyskens2015} who took into account the specific chemical composition of S stars and its feedback on the model-atmosphere structure. A new grid of \marcs\ model atmospheres dedicated to S stars \citep{VanEck2011} has been used for that purpose. \citet{Neyskens2015} obtained  $\Teff=3600$~K, $\log{g}=1.0$, \FeH$=-0.5$, C/O$=0.5$ and \sFe$=+1.0$. These parameters will be adopted in the present study.

For IT~Vir, a border case between Ba  and S stars, \citet{Abia98} estimated the atmospheric parameters to be $\Teff=4000$~K, $\log{g}=1.0$, and $[\mathrm{Fe/H}]=-1.0$ with $\xi=2$~\kms\ and  C/O$=0.95$, while \citet{Smith84} had previously found $\Teff=4200$~K, $\log{g}=0.8$, $[\mathrm{Fe/H}]=0.05$, and C/O$=0.79$. The $V-K$ index ($V-K=3.65$) combined with the calibration of \citet{Bessell-98} yields $T_{\rm eff} = 3900$~K. We adopted that value for the temperature combined with  \citet{Abia98} gravity.

The resulting atmospheric parameters are shown in Table~\ref{tab:sp}, sorted by increasing effective temperatures (as in all the following tables).
The uncertainties on the atmospheric parameters provided by the \bacchus\ pipeline are those resulting from the propagation of the line-to-line Fe-abundance scatter on the trends used to derive the atmospheric parameters, as explained above. 
Uncertainties on $\Teff$ lower than 50~K were arbitrarily set to a minimum value of 50~K.  

\begin{table*}
 \caption{Spectroscopic stellar parameters derived with \bacchus. Stars are ordered according to increasing $T_{\rm eff}$.}
  \center
 \begin{tabular}{lllcrcl}
   Star name      & HD       &     ~~~$\Teff$ (K) &      $\log{g}$  &     [Fe/H]~~~     &    $\xi$ (km~s$^{-1}$) \\
   \hline \\
 V1261 Ori\tablefootmark{a}   & 35155 &  $3600 \pm 100$ & $ 1.0 \pm 0.5$ & $-0.5  \pm 0.5~ $ &  $2.0  \pm 0.5 $ \\
 IT Vir\tablefootmark{a}      & 121447   &  $3900 \pm 100$ & $1.0 \pm 0.5 $& $-0.5 \pm 0.5$ & $2^b$ \\
 56 Peg      & 218356   & $4244 \pm 137$ & $ 1.3 \pm 0.4$ & $-0.45 \pm 0.12$ &  $1.55 \pm 0.06$ \\
 DR Dra      & 160538   & $4851 \pm  50$ & $ 3.0 \pm 0.5$ & $-0.14 \pm 0.10$ & $1.76 \pm 0.05$ \\
 $\zeta$ Cyg & 202109   & $4950 \pm  64$ & $ 2.7 \pm 0.1$ & $-0.03 \pm 0.08$ & $1.43 \pm 0.03$ \\
 IP Eri      &  18131   & $4958 \pm  50$ & $ 3.3 \pm 0.3$ & $ 0.09 \pm 0.08$ &  $1.45 \pm 0.04$ \\ 
 V1379 Aql   & 185510   & $5011 \pm 164$ & $ 3.2 \pm 0.2$ & $-0.23 \pm 0.12$ &  $1.68 \pm 0.17$ \\
 HR 5692     & 136138   & $5032 \pm  58$ & $ 3.0 \pm 0.2$ & $-0.08 \pm 0.09$ & $1.35 \pm 0.04$ \\
 $\zeta$ Cap & 204075   & $5329 \pm 101$ & $ 2.0 \pm 0.8$ & $-0.08 \pm 0.13$ &  $2.26 \pm 0.05$ \\ 
 63 Eri      &  32008   & $5374 \pm  50$ & $ 3.6 \pm 0.2$ & $-0.18 \pm 0.08$ &  $1.03 \pm 0.04$ \\  
 14 Aur C    &  33959 C & $6681 \pm  61$ & $ 4.1 \pm 0.3$ & $-0.11 \pm 0.08$ &  $1.75 \pm 0.09$ \\
 \hline

 \end{tabular}
 \tablefoot{
 \tablefoottext{a}{Atmospheric parameters not determined with BACCHUS}
\tablefoottext{b}{No uncertainty given by \citet{Abia98} from whom this value has been adopted}
 }
 \label{tab:sp}
\end{table*}

\section{Abundances}
\label{Sect:abundances}
We have chosen to perform the abundance analysis using spectrum synthesis. The detailed abundance analysis was performed using the \bacchus\ pipeline abundance module.  Only the least blended lines were retained for the analysis. This selection was performed over the whole HERMES wavelength range  (i.e., $[370-890]$~nm). The atomic line list (Table~\ref{Tab:all}) includes the isotopic shifts for \ion{Ba}{ii} (with an update for isotopes 130 and 132) and the hyperfine structure for \ion{La}{ii} from \citet{Masseron06}. The CH molecular line list is from \citet{Masseron14} and CN is from \citet{Sneden2014}. The references for the other molecular line lists (TiO, SiO, VO, C$_2$, NH, OH, MgH, SiH, CaH and FeH) can be found in \citet{Gustafsson2008}. Line fitting is based on a least-square minimization method and all lines are visually inspected to check for possible issues (caused by, e.g., line blends, cosmic hits, ...). Results are presented in Table~\ref{tab:cno_abu} for CNO abundances, and in Tables~\ref{tab:s_abu} and \ref{tab:sfe_abu} for Sr, Y, Zr, Ba, La, and Ce. The uncertainties given in Tables~\ref{tab:cno_abu} and \ref{tab:s_abu} correspond to the standard deviation of the line-to-line dispersion, whereas in Table~\ref{tab:sfe_abu}, the total uncertainties (thus including the impact of the uncertainties from the model parameters) are listed. 

\begin{table*}
  \caption{C, N, and O abundances. Only the statistical uncertainties are given here. The number of lines used are between parentheses. }
  \begin{center}
   \begin{tabular}{lrrrrc}
Star       &    [Fe/H]~~~~~&      [C/Fe]~~~~~~    & [N/Fe]~~~~~& [O/Fe]~~~~~& C/O \\
\hline\\
V1261 Ori\tablefootmark{a} & $-0.66\pm 0.09$ & $0.39\pm0.14$\tablefootmark{1}~~~~~~& $ 0.73\pm0.11$\tablefootmark{2}~~~~~~& $0.54\pm0.08$\tablefootmark{3}~~~~& 0.38\\ 
IT Vir\tablefootmark{b}     & $-0.45\pm0.25$ & $ 0.55~~~~~~~~~~~~~~~~~~~$ & $ 1.39\pm0.30~(50)$ & $ 0.40~~~~~~~~~~~~(1)$ & 0.75\\
           &                & $ 0.62~~~~~~~~~~~~~~~~~~~$ & $ 1.11\pm0.20~(94)$ & $ 0.60~~~~~~~~~~~~(1)$ & 0.90\\  
56 Peg     & $-0.45\pm0.12$ & $ 0.01\pm0.23~(13)$ & $-0.17\pm0.07~(82)$ & $ 0.12\pm~~-~~~(1)$ & 0.63\\
DR Dra     & $-0.14\pm0.10$ & $-0.23\pm0.07~~~(8)$& $-0.21\pm0.09~(82)$ & $ 0.15\pm~~-~~~(1)$ & 0.39\\
$\zeta$ Cyg& $-0.03\pm0.08$ & $-0.22\pm0.13~(15)$ & $ 0.11\pm0.05~(94)$ & $ 0.16\pm0.06~(2)$  & 0.44\\
IP Eri     & $ 0.09\pm0.08$ & $-0.14\pm0.16~(18)$ & $-0.23\pm0.06~(96)$ & $ 0.14\pm~~-~~~(1)$ & 0.54\\
V1379 Aql  & $-0.23\pm0.12$ & $-0.24\pm0.13~~~(8)$& $ 0.21\pm0.05~(29)$ &     \multicolumn{1}{c}{$-$} &    \multicolumn{1}{c}{$-$}  \\
HR 5692    & $-0.08\pm0.09$ & $-0.24\pm0.10~(10)$ & $-0.16\pm0.08~(90)$ & $ 0.20\pm~~-~~~(1)$ & 0.43\\ 
$\zeta$ Cap& $-0.08\pm0.13$ & $ 0.11\pm0.05~(16)$ & $ 0.38\pm0.07~(85)$ & $ 0.12\pm0.08~(2)$  & 0.79\\   
63 Eri     & $-0.18\pm0.08$ & $-0.12\pm0.18~(13)$ & $ 0.25\pm0.05~(50)$ & $ 0.31\pm0.04~(6)$  & 0.20\\ 
14 Aur C   & $-0.11\pm0.08$ & $ 0.22\pm0.03~~~(3)$& $ 0.25\pm0.13~~~(4)$ & $ 0.16\pm~~-~~~(1)$ & 0.72\\ 
\\ 
\hline
\end{tabular} 
\tablefoot{
 \tablefoottext{a}{For this star, we used the Fe abundance from \citet{Kovacs1983} scaled to the solar value of \citet{Holweger1979}. CNO abundances are from \citet{Smith1990} and are relative to $\alpha$~Tau }
\tablefoottext{b}{Two equally-likely abundance sets are listed for IT~Vir (see text).}
 \tablefoottext{1}{from CO lines at 1.6 and 2.2 $\mu$m }
 \tablefoottext{2}{from CN lines at 2.2 $\mu$m }
 \tablefoottext{3}{from OH lines at 1.6 and 2.2 $\mu$m }
 }

  \end{center}
 \label{tab:cno_abu}
\end{table*}
\begin{table*}
  \caption{Absolute s-process element abundances $A(\text{X})$ (defined as $A(\text{X}) = \log{[n(\text{X})/n(\text{H})]} + 12$, where $n(\text{X})$ and $n(\text{H})$ are the number densities of X and H). Only the statistical uncertainties are given here. The number of lines used are between parentheses.}
  \begin{center}
   \begin{tabular}{lrrrrrr}
Star       & \multicolumn{1}{c}{$A($Sr$)$} & \multicolumn{1}{c}{$A($Y$)$}& \multicolumn{1}{c}{$A($Zr$)$} & \multicolumn{1}{c}{$A($Ba$)$} & \multicolumn{1}{c}{$A($La$)$}& \multicolumn{1}{c}{$A($Ce$)$} \\
\hline
\\
V1261 Ori  & \multicolumn{1}{c}{$-$} & $2.66\pm0.27$~~~(3) & $3.48\pm0.17$~~~(4) & $3.02\pm0.32$~~~(3) & $1.91\pm0.06$~~~(2) & $2.33\pm0.07$~~~(7)\\
IT Vir     & $3.44\pm~~-~~$~~(1) & $2.70\pm0.12$~~~(3) & $3.99\pm0.26$~~~(7) & $4.05\pm0.18$~~~(3) & $2.42\pm0.23$~~~(4) & $2.52\pm0.20$~~~(8)\\
56 Peg     & $3.05\pm0.06$   (2) & $2.22\pm0.22$~~~(8) & $2.44\pm0.07$~~~(7) & $3.18\pm0.18$~~~(3) & $1.12\pm0.09$~~~(8) & $1.41\pm0.09$~~~(6)\\
DR Dra     & $3.04\pm0.05$   (2) & $2.13\pm0.22$~~~(5) & $2.77\pm0.06$~~~(5) & $2.46\pm0.08$~~~(4) & $1.17\pm0.10$~~~(8) & $1.56\pm0.07$~~~(4)\\
$\zeta$ Cyg& $3.33\pm~~-~~$~~(1) & $2.61\pm0.16$~~~(7) & $2.96\pm0.20$~~~(9) & $3.16\pm0.25$~~~(4) & $1.52\pm0.08$~~~(9) & $1.87\pm0.12$  (11)\\
IP Eri     & $3.14\pm~~-~~$~~(1) & $2.06\pm0.12$~~~(9) & $2.86\pm0.20$~~~(7) & $2.53\pm0.09$~~~(3) & $1.25\pm0.17$~~~(6) & $1.60\pm0.09$~~~(4)\\
V1379 Aql  & $2.89\pm~~-~~$~~(1) & $2.23\pm0.04$~~~(4) & $2.59\pm0.02$~~~(2) & $2.29\pm0.07$~~~(3) & $1.11\pm0.05$~~~(3) & $1.52\pm0.16$~~~(3)\\
HR5692     & $3.16\pm~~-~~$~~(1) & $2.63\pm0.20$~~~(6) & $3.14\pm0.19$  (11) & $2.90\pm0.07$~~~(5) & $1.49\pm0.10$  (11) & $1.77\pm0.13$~~~(9)\\
$\zeta$ Cap& $3.86\pm0.06$   (2) & $3.50\pm0.21$~~~(9) & $3.87\pm0.28$  (11) & $4.06\pm0.13$~~~(3) & $2.09\pm0.10$  (11) & $2.47\pm0.15$  (10)\\
63 Eri     & $2.84\pm0.01$   (2) & $2.06\pm0.13$  (10) & $2.48\pm0.17$~~~(8) & $2.64\pm0.22$~~~(4) & $1.15\pm0.06$  (10) & $1.55\pm0.06$~~~(9)\\
14 Aur C   & $3.04\pm0.05$   (3) & $2.03\pm0.12$~~~(7) & $2.64\pm0.07$~~~(5) & $2.60\pm0.16$~~~(4) & $1.22\pm0.11$~~~(5) & $1.50\pm0.20$~~~(3)\\
\\
\hline
Sun\tablefootmark{a} & $2.92\pm0.05$~~~ & $2.21\pm0.02$~~~& $2.58\pm0.02$~~~ & $2.17\pm0.07$~~~& $1.13\pm0.05$~~~ & $1.70\pm0.10$~~~ \\
\hline
\end{tabular} 
\tablefoot{
 \tablefoottext{a}{Solar abundances from \citet{Grevesse07}.}
 }

  \end{center}
 \label{tab:s_abu}
\end{table*}

\subsection{C, N, O abundances} 

C, N, and O abundances were derived in the following sequence: first O, then C and finally N.
For deriving the oxygen abundance, we mainly used the [\ion{O}{i}] 630.030 and  636.378~nm  lines since the \ion{O}{i} triplet resonance line at 777 nm is strongly affected by NLTE effects [see \citet{Asplund2005} for a detailed discussion]. The \ion{O}{i} triplet  line yields a LTE abundance larger by about $\sim0.2-0.4$~dex as compared to the abundance derived from the forbidden lines. This is not the case for the star 63~Eri, however, since for that star, the triplet and the strong doublet at 844.636 and 844.676~nm give consistent abundances within 0.04~dex, even though the [\ion{O}{i}] 630.030~nm is blended with a telluric line in its blue wing (which has not been considered in the fit procedure thus). This star has one of the largest gravity, leading to a decrease of the non-LTE effects on the 777~nm triplet. For 14~Aur~C, the two forbidden lines are lost in the noise, so that we use instead the highly excited, very weak line at 615.68~nm.

Although it is particularly difficult to find unblended atomic lines of \ion{C}{i}, we found 5 useful weak lines (as listed in Table~\ref{Tab:all}), except in V1379~Aql, V1261~Ori and IT~Vir where they are not visible or too blended to be useful. This is why we prefer to use lines of diatomic carbon C$_2$ for deriving carbon abundances. Lines of C$_2$ appear roughly between 400 and 600~nm. In the molecular linelists, we first identified useful lines of C$_2$ by comparing synthetic spectra (for typical values of the atmospheric parameters) with only C$_2$ lines included and with all lines included to identify the least blended C$_2$ lines. Then by comparing with observed spectra, we ended up with a selection of $\sim$20 lines (Table~\ref{Tab:mll}) visible in at least two-thirds of the objects. The line at 513.56~nm used by \citet{Barbuy1992} is also included. Good fits of C$_2$ bands are difficult to achieve 
but the lines are so sensitive to the carbon abundance that the latter can be easily inferred  even from rough fits. In 14 Aur~C, rotation broadens the lines so much (see Fig.~\ref{fig:spec_baii}) that only three lines, despite being blended with others, can nevertheless be used to derive 
the carbon abundance. Especially the C$_2$ line at 468.427~nm appears useful for this star (it is not used for the other stars because the line position in the synthetic spectrum appears to be somewhat redder than observed). 
The C/O ratio is then calculated and listed in Table~\ref{tab:cno_abu}.

We selected ten \ion{N}{i} lines (Table~\ref{Tab:all}).
But for the range of atmospheric parameters under consideration, 
\ion{N}{i} lines are always very weak and blended with CN lines. The N abundance thus  combines \ion{N}{i} and CN lines, using the carbon abundance from the previous step. The CN lines are numerous and span the full HERMES spectral range. 
We selected 121 CN lines unblended or weakly blended, in the range 640 to 890~nm (Table~\ref{Tab:mll}). All CN lines are eye-checked because a lot of telluric lines are present in this wavelength range and can strongly blend with CN lines. Below 500~nm, CN lines are too much blended by atomic and molecular features (mainly from CH, C$_2$, MgH, and SiH) to be of any use. NH lines cannot be used because they are located beyond the violet edge of the spectrograph (370~nm). For the fast rotator 14~Aur~C, almost no CN lines can be used, since weak lines blend together due to rotational broadening.  We nevertheless used 2 lines among the selected CN lines, plus 2 other weak lines not in the original selection (at $\sim$~871.15 and 871.88~nm). The N abundance in 14~Aur~C is uncertain since the weak lines are at the limit of the spectral noise.

The S star V1261~Ori requires a specific way to derive its carbon abundance since atomic and C$_2$ lines are flooded in TiO and ZrO bands.
Infrared bands from CO, CN and OH around 2~$\mu$m must be used instead, as done by \citet{Smith1990} (with atmospheric parameters $T_{\rm eff} = 3650$~K, $\log g = 0.8$, [Fe/H] = $-0.52$). We therefore used their CNO values for determining the subsequent s-process elements.

The derivation of CNO abundances in IT~Vir is particularly challenging and consistently leads to a [N/Fe] ratio above 1.0~dex. 
We proceeded as follows: we first estimated the O abundance through the [\ion{O}{I}] 630.0 nm line using spectral syntheses computed from dedicated S-star model atmospheres with the parameters listed in Table~\ref{tab:sp}, and three different values of the C/O ratio (0.50, 0.75 and 0.90). Basically, as the C/O ratio increases (for a given O abundance), the strength of the forbidden line decreases. This strength is well reproduced for [O/Fe] = $+0.20$, $+0.40$ and $+0.60$~dex, respectively. 
Then, for each C/O ratio and the corresponding O abundance, a carbon abundance ensues. We checked which one among these carbon abundances best fit the 
C$_2$ lines (between 460 and 560~nm) but we could only exclude the value corresponding to C/O = 0.5; the other two possibilities leading to similarly good fits, they could not be discriminated.  
From the above analysis, we can at least be sure that the C/O ratio is intermediate between the solar value and unity, thus disqualifying the SC spectral type quoted in the literature. 
For each remaining pair of (C/O, [O/Fe]) values, we finally compute synthetic spectra for three different  [N/Fe] values  ($0.0$, $+1.0$ and $+1.5$~dex)
and evaluate how well these synthetic spectra fit the observed spectrum in the three wavelength ranges [629 -- 631], [776.5 -- 778.5], and [843.5 -- 845.5]~nm containing CN lines. Good fits are obtained for [N/Fe$]=1.5$~dex when C/O$=0.75$, [O/Fe$]=0.40$~dex and for [N/Fe$]=1.0$~dex when C/O$=0.90$, [O/Fe$]=0.60$. Again, we cannot discriminate between these two equally good fits. Therefore both possibilities are listed in Table~\ref{tab:cno_abu}. Nevertheless, whenever necessary in the remainder of the study, we will adopt the former set.

We checked the sensitivity to gravity $\log g$ of our conclusion of a large N overabundance. Adopting $\log g = 0$ instead of 1 as above, all other parameters being equal (with C/O$=0.75$), we  obtain a value of [N/Fe] still larger than 1.0.
Similarly, increasing the temperature does not permit to
lower the [N/Fe] value below 1.0~dex. We believe therefore that this result is robust.

\begin{table*}
  \caption{Abundances of s-process elements. Total uncertainties are given here.}
  \begin{center}
   \small
\begin{tabular}{lrrrrrr|r}
Star       & [Sr/Fe]~~~~~~~~& [Y/Fe]~~~~~~~~& [Zr/Fe]~~~~~~~~ & [Ba/Fe]~~~~~~~~ & [La/Fe]~~~~~~~~& [Ce/Fe]~~~~~~~~& [s/Fe]~~~~ \\
\hline
\\
V1261 Ori  & \multicolumn{1}{c}{-}       & $ 1.11\pm0.58$~~~(3) & $1.56\pm0.54$~~~(4) & $1.51\pm0.62$~~~(3) & $1.44\pm0.54$~~~(2) & $ 1.29\pm0.24$~~~(7) &  $1.38\pm0.18$ \\
IT~Vir     & $0.97\pm0.26$  (1) & $ 0.94\pm0.30$~~~(3) & $1.86\pm0.38$~~~(7) & $2.33\pm0.35$~~~(3) & $1.74\pm0.39$~~~(4) & $ 1.27\pm0.37$~~~(8) &  $1.52\pm0.55$ \\
56 Peg     & $0.58\pm0.16$  (2) & $ 0.46\pm0.28$~~~(8) & $0.31\pm0.17$~~~(7) & $1.46\pm0.28$~~~(3) & $0.44\pm0.24$~~~(8) & $ 0.16\pm0.24$~~~(6) &  $0.57\pm0.46$ \\
DR Dra     & $0.26\pm0.14$  (2) & $ 0.60\pm0.27$~~~(5) & $0.33\pm0.16$~~~(5) & $0.43\pm0.22$~~~(4) & $0.18\pm0.23$~~~(8) & $ 0.00\pm0.22$~~~(4) &  $0.21\pm0.16$ \\
$\zeta$ Cyg& $0.34\pm0.12$  (1) & $ 0.42\pm0.22$~~~(7) & $0.39\pm0.24$~~~(9) & $0.89\pm0.31$~~~(4) & $0.40\pm0.22$~~~(9) & $ 0.21\pm0.24$  (11) &  $0.49\pm0.28$ \\
IP Eri     & $0.13\pm0.12$  (1) & $-0.24\pm0.19$~~~(9) & $0.19\pm0.24$~~~(7) & $0.27\pm0.21$~~~(3) & $0.03\pm0.26$~~~(6) & $-0.19\pm0.22$~~~(4) &  $0.03\pm0.21$ \\
V1379 Aql  & $0.20\pm0.15$  (1) & $ 0.25\pm0.18$~~~(4) & $0.24\pm0.16$~~~(2) & $0.35\pm0.22$~~~(3) & $0.21\pm0.23$~~~(3) & $ 0.05\pm0.27$~~~(3) &  $0.22\pm0.10$ \\
HR5692     & $0.32\pm0.12$  (1) & $ 0.50\pm0.25$~~~(6) & $0.64\pm0.23$  (11) & $0.81\pm0.21$~~~(5) & $0.44\pm0.23$  (11) & $ 0.15\pm0.24$~~~(9) &  $0.48\pm0.23$ \\
$\zeta$ Cap& $1.02\pm0.17$  (2) & $ 1.37\pm0.28$~~~(9) & $1.37\pm0.33$  (11) & $1.97\pm0.25$~~~(3) & $1.04\pm0.25$  (11) & $ 0.85\pm0.27$  (10) &  $1.27\pm0.40$ \\
63 Eri     & $0.10\pm0.12$  (2) & $ 0.03\pm0.20$  (10) & $0.08\pm0.22$~~~(8) & $0.65\pm0.29$~~~(4) & $0.20\pm0.21$  (10) & $ 0.03\pm0.21$~~~(9) &  $0.18\pm0.24$ \\
14 Aur C   & $0.23\pm0.13$  (3) & $-0.07\pm0.19$~~~(7) & $0.17\pm0.15$~~~(5) & $0.54\pm0.25$~~~(4) & $0.20\pm0.23$~~~(5) & $-0.09\pm0.29$~~~(3) &  $0.16\pm0.23$ \\
\\
\hline
\end{tabular}
\normalsize
  \end{center}
 \label{tab:sfe_abu}
\end{table*}

\subsection{\textit{s}-process abundances}
Atomic lines were carefully selected over the whole HERMES spectral range using both synthetic  and observed spectra. They are listed in Table~\ref{Tab:all}. Whenever possible, for each species, we tried to select lines from both neutral and singly-ionized states (excluding strong resonance lines of neutral species known to suffer from  NLTE effects).

The only observable lines of \ion{Sr}{ii} are the strong ones below 430.5~nm which are too strongly blended to derive reliable abundances.
We therefore only used \ion{Sr}{i} lines.
Among the 8 selected \ion{Sr}{i} lines, the line at 687.83~nm is often blended by telluric O$_2$,  the resonance line at 460.73~nm  is known to suffer from NLTE effects and to give abundances lower than that from higher excitation lines.
The line at 707.01~nm has been used in all the target stars but 14~Aur~C and V1261~Aur. 
In 14~Aur~C, the two lines at 640.85 and 707.01~nm are not observable because they blend with other lines due to rapid rotation or are lost in the noise. Instead, the strong \ion{Sr}{ii} line  around 416~nm seems to provide a reliable estimate of the Sr abundance. 
In V1261~Ori, the two \ion{Sr}{i} lines at 640.85 and 707.01~nm are too blended and we used instead two \ion{Sr}{i} lines at 687.83 and 689.26~nm even though the continuum placement remains the largest uncertainty.

\begin{figure*}
 \includegraphics[angle=-90, width=1.\linewidth]{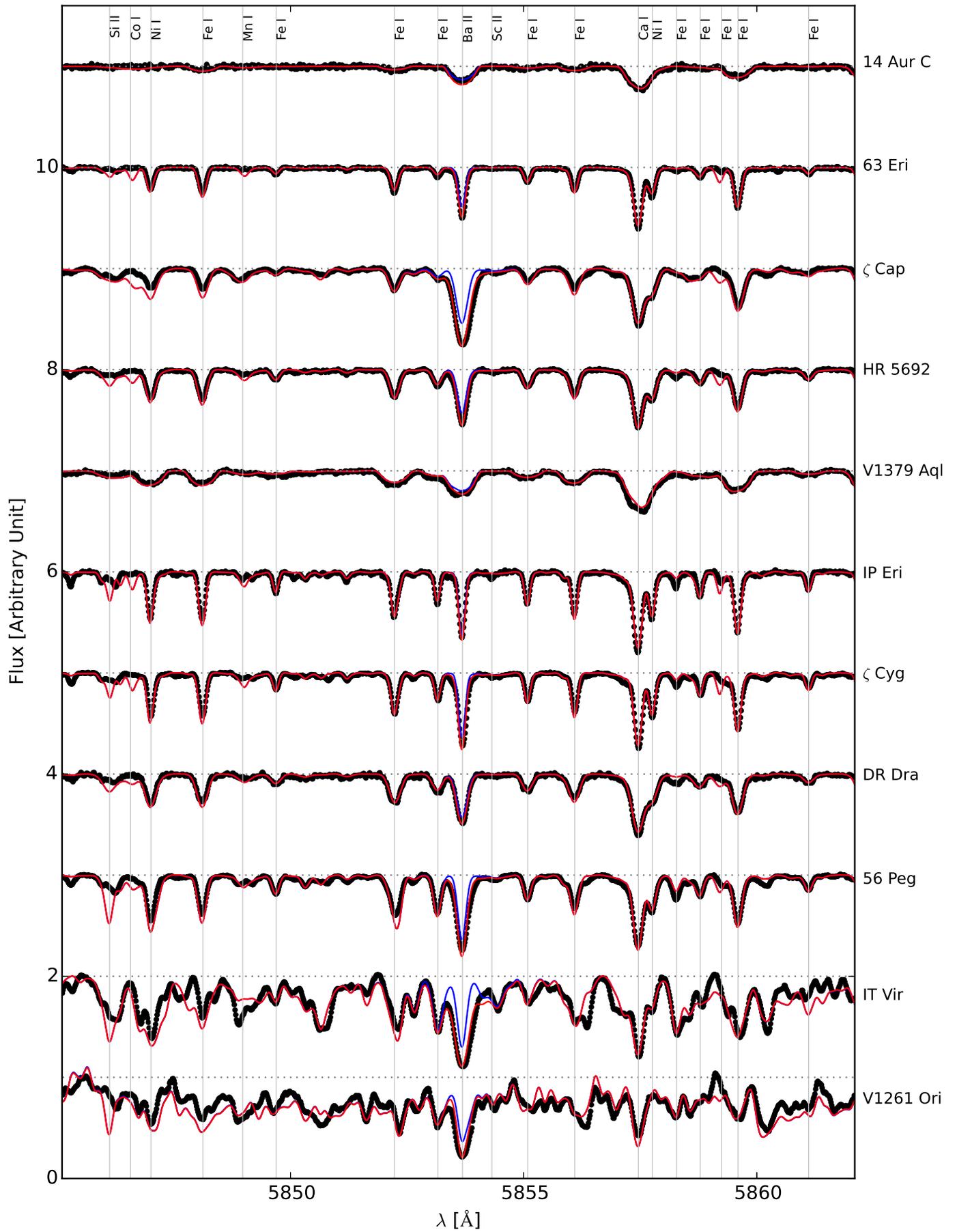}
 \caption{Example of the observed and synthetic spectra around the \ion{Ba}{II} line at 585.36 nm. The solid blue line corresponds to [Ba/Fe] = 0.}
 \label{fig:spec_baii}
\end{figure*}

For Y, we mainly used clean lines of \ion{Y}{ii} that are more numerous than \ion{Y}{i} lines. 
The cleanest and most useful \ion{Y}{ii} lines in our sample are located at 528.98, 540.28, 572.89, and 679.54~nm. The other lines used are listed in Appendix~\ref{ap:ll}.

For Zr, we mainly used lines of \ion{Zr}{i}, since they are more numerous than \ion{Zr}{ii} lines. 
Lines of \ion{Zr}{i} between 800 and 850~nm, when useful, give abundances systematically stronger, by about $\sim0.5$~dex. For 14~Aur~C, only \ion{Zr}{ii} lines below 540~nm can be used for the abundance determination. Since V1261~Ori is veiled by ZrO molecular bands, we used the same lines (between 740 and 758~nm) as those used by \citet{Smith1990}  to derive the Zr abundance. Lines above 800~nm are useable but give somewhat higher abundances, as already mentioned.

For s-process elements belonging to the second peak  (Ba, La and Ce), only neutral lines are available for Ba. We do not use  the \ion{Ba}{ii} resonance lines at 455.4 and 493.4~nm in the abundance analysis due to known NLTE effects (see e.g., \citealt{Olshevsky2007}), except in 14~Aur~C where too few other lines are available.
We mainly use the \ion{Ba}{ii} lines at 585.37, 614.17, 649.69~nm and the \ion{Ba}{i} line at 748.81~nm.

Many  lines (about 30) are available for La and Ce in the HERMES wavelength range. 
 The most useful \ion{La}{II} lines are located at 529.08,  530.20, 530.35, 617.27, 639.05, and  677.43~nm. The most useful \ion{Ce}{II} lines are located at 456.24, 533.06, 604.34, 802.56, 871.67, and 877.21~nm. The \ion{Ce}{II} lines are the  strongest in V1261~Ori, which indeed has the largest Ce overabundance ([Ce/Fe] = 1.29~dex) in our sample. 

To illustrate the good agreement between observed and synthetic spectra, we show in Fig.~\ref{fig:spec_baii}  the spectral region around  the \ion{Ba}{II} 585.36~nm line, which is well visible in the entire sample. The synthetic spectra in blue are computed with [Ba/Fe$]=0$ whereas the red ones are computed with the appropriate [Ba/Fe] abundances as listed  in Table~\ref{tab:sfe_abu}. The red synthetic spectra reproduce nicely the observed spectra, especially for the fast rotators V1379~Aql and 14~Aur~C.  The fit is a bit worse for the cooler S stars V1261~Ori and IT~Vir because of the presence of numerous molecular lines, but it is nevertheless possible to derive reliably the Ba abundance.

\subsection{Uncertainties on \textit{s}-process abundances}
We investigated the impact on the s-process abundances of the uncertainties from the atmospheric parameters, adopting as typical uncertainties
on these parameters (Table~\ref{tab:sp}) the values  $\sigma_{\mathrm{T}_{\mathrm{eff}}}= \pm 50$~K, $\sigma_{\log{g}}= \pm 0.3$~dex, $\sigma_\mathrm{[Fe/H]}=\pm 0.10$~dex  and $\sigma_\xi= \pm 0.10$~km~s$^{-1}$. 

The statistical uncertainty $\sigma_\mathrm{stat}$ listed in Table~\ref{tab:err} is  the line-to-line dispersion divided by $\sqrt{N}$, where $N$ is the number of lines used for a given element (Table~\ref{tab:err}). This statistical uncertainty takes into account not only uncertainties on atomic data but also on continuum location and line-profile mismatch issues.

Statistical ($\sigma_\mathrm{stat}$) and model-induced (i.e., $\sigma_{\mathrm{T}_{\mathrm{eff}}}$, $\sigma_{\log{g}}$, $\sigma_\mathrm{[Fe/H]}$ and $\sigma_\xi$) uncertainties are quadratically added \citep[thus assuming independence between them, as is common practice, although in reality they may be correlated, as discussed by][]{2002ApJS..139..219J} to obtain the total uncertainty $\sigma_\mathrm{tot}$ for each \textit{s}-process element 
(which may thus be overestimated).  We perform this analysis on   $\zeta$~Cyg as it is representative of the whole sample (see Fig.~\ref{fig:ap}). The estimated uncertainties are given in Table~\ref{tab:err}. Note that the uncertainties on the absolute s-process abundances  given in Table~\ref{tab:s_abu} are only the line-by-line dispersion. 

\begin{table}
 \caption{The uncertainties on the derived abundances in $\zeta$~Cyg: $\sigma_{i}$ are the uncertainties caused by the errors on the atmospheric parameters ($i = T_{\mathrm{eff}}$,  $\log{g}$, $\mathrm{[Fe/H]}$, $\xi$; see text) and $\sigma_\mathrm{stat}$ is the line-to-line scatter on the abundance divided by $\sqrt N$, where $N$ is the number of lines used. The individual model-driven uncertainties are quadratically added to yield the 
total model-driven uncertainty $\sigma_\mathrm{atm}$, which in turn is quadratically added to the statistical uncertainty to give the total uncertainty $\sigma_\mathrm{tot}$.
}
 \small
 \begin{tabular}{lrcccccc|cc}
\hline\\
 X & $N$ &  $\sigma_{\mathrm{T}_{\mathrm{eff}}}$ & $\sigma_{\log{g}}$ & $\sigma_\mathrm{[Fe/H]}$  &$\sigma_\xi$ & $\sigma_\mathrm{atm}$  & $\sigma_\mathrm{stat}$ & $\sigma_\mathrm{tot}$\\
 \hline\\
 Sr &  1 & 0.05 & 0.01 & 0.01 & 0.01 & 0.05 &   -  & 0.05\\
 Y  &  7 & 0.03 & 0.10 & 0.01 & 0.04 & 0.11 & 0.06 & 0.13\\
 Zr &  9 & 0.09 & 0.01 & 0.01 & 0.01 & 0.09 & 0.07 & 0.12\\
 Ba &  4 & 0.09 & 0.09 & 0.06 & 0.05 & 0.15 & 0.11 & 0.19\\
 La &  9 & 0.03 & 0.16 & 0.03 & 0.01 & 0.17 & 0.03 & 0.17\\
 Ce & 11 & 0.03 & 0.14 & 0.05 & 0.02 & 0.15 & 0.04 & 0.16\\
 \\
 \hline
 \end{tabular}
  \normalsize
  \label{tab:err}
\end{table}  %

\begin{figure}
 \includegraphics[width=\linewidth]{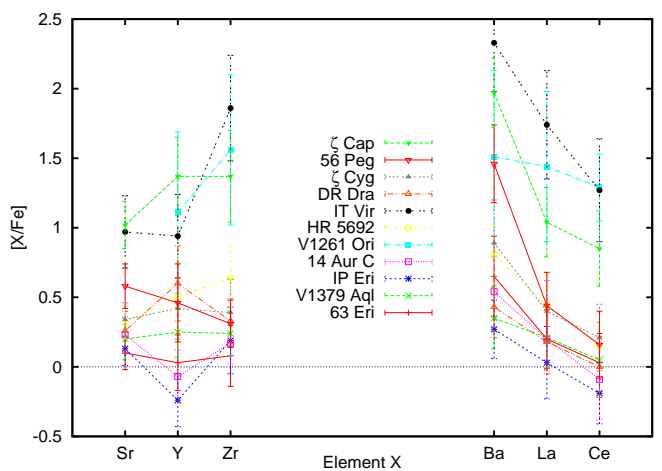}
 \caption{First-peak (Sr, Y, Zr) and second-peak (Ba, La, Ce) s-process abundances relative to Fe.}
 \label{fig:s_abu_z}
\end{figure}

\begin{figure*}
\includegraphics[width=0.9\linewidth]{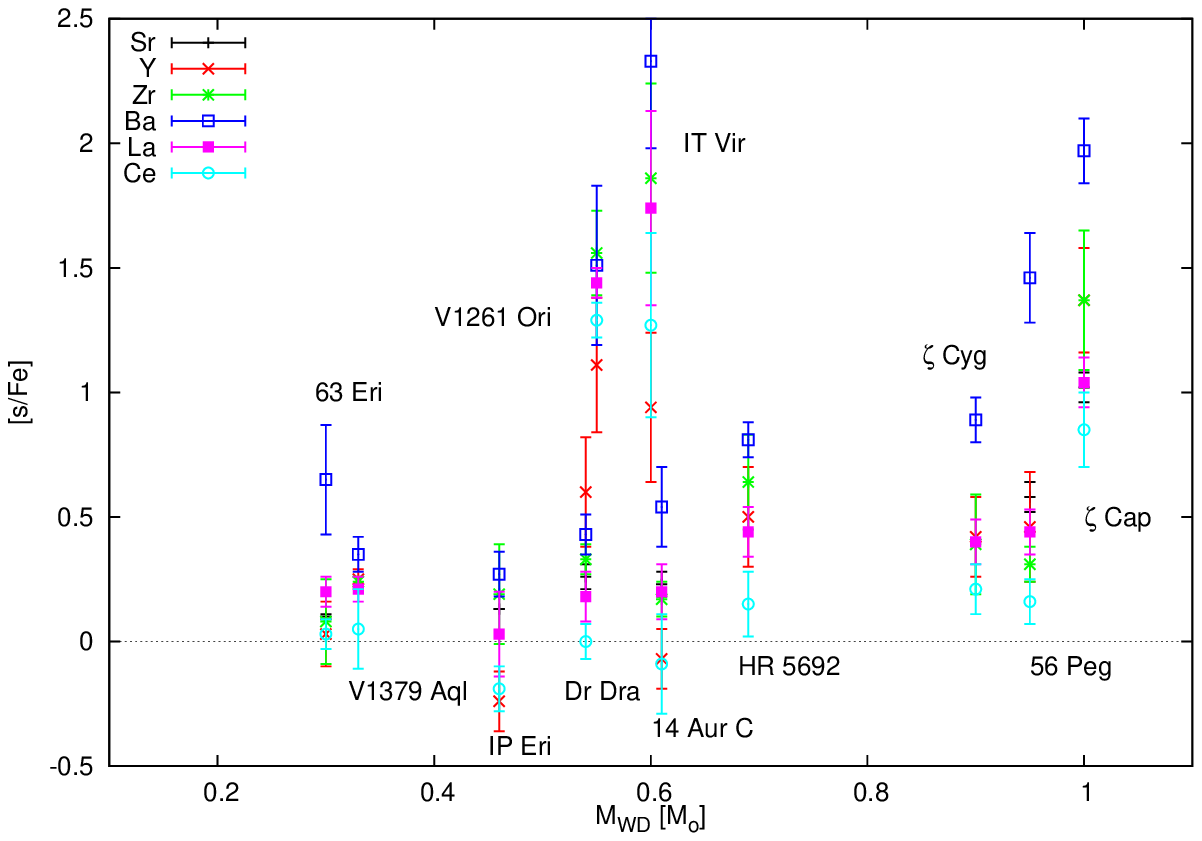}

 \includegraphics[width=0.9\linewidth]{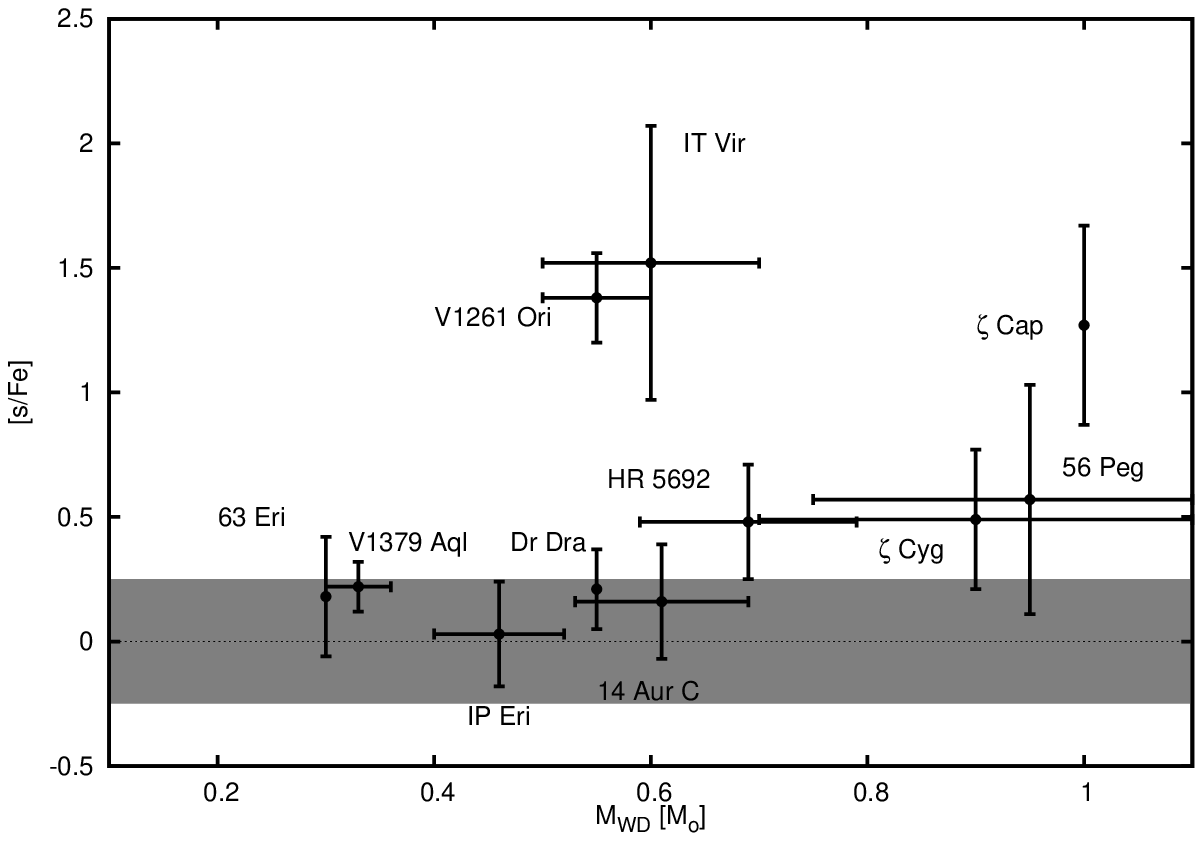}
 \caption{Upper panel: First (Sr, Y, Zr) and second peak (Ba, La, Ce) s-process abundances relative to Fe as a function of the WD mass (a color version is available online). Lower panel: Same as left panel but with mean s-process abundances relative to Fe. To guide the eye, the grey zone marks the $\pm0.25$~dex zone with no significant enhancement in s-process abundances.}
 \label{fig:s_abu}
\end{figure*}

To get the uncertainties on the various [X/Fe] abundances (as listed in  Table~\ref{tab:sfe_abu}), we add quadratically the uncertainties on the absolute abundances $A$(Fe) and $A$(X) (as defined in footnote 1). The resulting uncertainties are expressed as:
\begin{equation}
\sigma_\mathrm{X} = \sqrt{\sigma_\mathrm{tot}(\mathrm{X})^2 + \sigma(\mathrm{Fe})^2 + \sigma_\odot(\mathrm{X})^2 + \sigma_\odot(\mathrm{Fe})^2}
\end{equation}
where $\sigma_\odot(\mathrm{X})$ denotes the uncertainty on the solar abundance of species $X$ as given by \citet{Grevesse07}. The corresponding  $\sigma_\mathrm{X}$ uncertainties are listed in Table~\ref{tab:sfe_abu} and plotted on Figs.~ \ref{fig:s_abu_z} and \ref{fig:s_abu}.

\subsection{Comparison with previous studies}

\noindent{\bf V1261~Ori (HD~35155).} That star has been analysed by \citet{Smith1990}, who find the second-peak s-process elements (Ba, La, Ce, Nd) to be enriched by more than 1~dex, as we do. However, 
\citet{Smith1990} find no overabundance in excess of 0.23~dex for Y and Zr, whereas we find these elements to be overabundant by more than 1~dex. We stress that both the models and the line data are very different between the two studies.  

\noindent {\bf IT Vir.} \citet{Smith84} derived the CNO abundances in IT~Vir, but does not confirm the large N abundance  that we obtain for that star ([N/Fe$]=1.1$ to 1.4~dex -- Table~\ref{tab:cno_abu}-- , as compared to 0.52~dex for Smith 1984). This large N overabundance goes along with a small $^{12}$C/$^{13}$C ratio of 8 \citep{Smith84}, which could hint at CN processing. However, similarly small  $^{12}$C/$^{13}$C ratios are a common feature among barium stars \citep{Smith84,Barbuy1992}, without being associated with N overabundances as large as the one we find for IT~Vir. 
We found that, whatever effective temperature or gravity is adopted for this star ($\Teff = 3900$ or $4200$~K, $\log g = 0$ or 1~dex; see Sect.~\ref{Sect:atmosphere}), the large [N/Fe] overabundance remains. Even after carefully re-deriving O and C abundances, [N/Fe] is still larger than 1~dex. Formerly, the largest N abundances observed in barium stars were [N/Fe]~=~0.7~dex in HD~27271 and 0.65~dex in HD~60197 \citep{Barbuy1992}. It may be relevant here to note that IT~Vir has one of the largest Roche-lobe filling factor among barium stars, since it has a moderately short orbital period (186~d), but a comparatively large radius \citep[$\sim 55$~R$_{\odot}$;][]{Jorissen1995}. As a consequence, these authors report ellipsoidal variations for that star. The associated faster rotation, tidal deformations and departure from spherical symmetry will trigger meridional currents and deep mixing, and the observed large N abundance in IT~Vir could perhaps be related to such  processes  currently occurring in the red giant \citep[see for instance][for a discussion of rotation-induced mixing, and the resulting surface enhancement in nitrogen]{Decressin2009}.

Regarding s-process abundances, the agreement is good for Sr, Y, and Ce, but Zr, Ba, and La are found to be more abundant by about 1~dex in our study as compared to \citet{Smith84}.

\noindent{\bf 56~Peg.} It is the first time that an s-process abundance analysis is performed for 56~Peg, and it confirms its barium nature. Because of 56~Peg high luminosity (it is classified as K0Iab -- confirmed by the low gravity of $\log g = 1.3$; Table~\ref{tab:sp}), one may have feared that the increased strength of the \ion{Ba}{II} lines, which led to its classification as a barium star, were caused by the low atmospheric pressure, a consequence of  Saha equation. But that fear is unjustified, since the present abundance analysis reveals a weak albeit definite s-process enhancement.

\noindent{\bf $\zeta$~Cap and $\zeta$~Cyg.} These stars were analysed by \citet{Smiljanic2007}. The agreement between the two studies on the s-process abundances is reasonable (in particular, both studies find large overabundances in   $\zeta$~Cap and much milder ones in $\zeta$~Cyg), except for Sr in $\zeta$~Cap: \citet{Smiljanic2007} obtain [Sr/Fe]~=~2.21, as compared to 1.02 in the present study. \citet{Smiljanic2007} used the single 
\ion{Sr}{i} resonance line at 460.73~nm, whereas we used that line and a line at 707.01~nm.  Based solely on the 460.73~nm line,  we find [Sr/Fe$]=0.99$, and that value is in good agreement with the value 1.07  based on the 707.01~nm line. The reason for the discrepancy between the Sr abundances derived from the \ion{Sr}{i} resonance line at 460.73~nm here and in \citet{Smiljanic2007} must partly reside in the different  $\log{gf}$ values used [0.069 by \citet{Smiljanic2007} against 0.283 in our work, from NIST with the AA accuracy -- see Appendix~\ref{ap:ll}], and partly in the different atmospheric parameters adopted.

\noindent{\bf DR~Dra.} \citet{1994AstL...20..796B}, \citet{1997A&AS..122...31Z}, and \citet{2010BaltA..19..157B} conclude, like we do, that DR~Dra does not exhibit any s-process overabundance.

\noindent{\bf IP~Eri.} That star has been analyzed by \citet{2014A&A...567A..30M}. We have rederived the s-process abundances with the same pipeline and methodology, but with a refined line selection for each element. The rederived s-process abundances agree with our previous work within the error bars and we obtain in the present study lower statistical uncertainties.

\begin{figure}
 \includegraphics[width=0.95\linewidth]{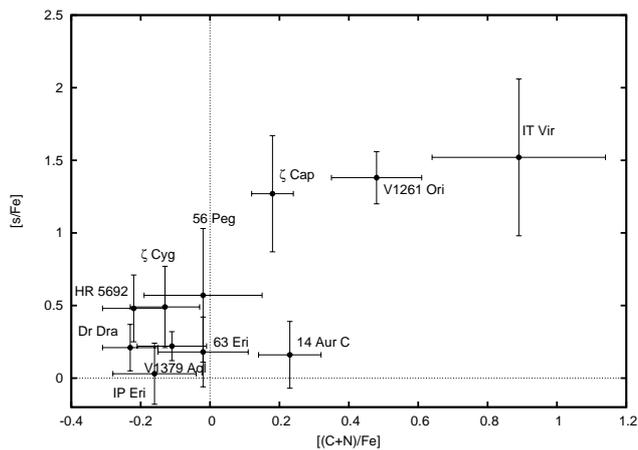}
 \caption{\label{fig:cfe_sfe}
The correlation between the s-process abundances and the 
carbon + nitrogen abundances.}
\end{figure}

\section{Discussion}
\label{Sect:discussion}

The s-process abundance distribution is presented in Fig.~\ref{fig:s_abu_z} where it is clearly apparent that our sample contains both normal stars and stars enriched in s-process elements (and the latter also exhibit enhanced carbon and sometimes nitrogen overabundances; Fig.~\ref{fig:cfe_sfe}). The origin of the difference between these two stellar categories is clearly revealed when plotting the average s-process abundance, as listed in the last column of Table~\ref{tab:sfe_abu}, as a function of the mass of the WD companion  (Fig.~\ref{fig:s_abu}).  The expected trend is indeed obtained: s-process overabundances are found in those binary stars with a companion WD mass larger than about 0.5~\Mo , i.e., the progenitor of the WD went through the thermally-pulsing AGB (TP-AGB) phase, was able to synthesize s-process elements in the intershell zone, and these were subsequently dredged up to the surface. Indeed, \citet{2000MNRAS.315..543H} (their Eq.~66) predict the minimum CO core mass at the base of the AGB
(just at the end of core He-burning) to be at least 0.51~\Mo\ (for a star of
initial mass 0.9~\Mo). Conversely, when the WD mass is smaller than 0.51~\Mo, the s-process synthesis did not occur (because the progenitor did not reach the TP-AGB): s-process overabundances are not observed. There are two marginal cases, namely DR~Dra and 14~Aur~C, whose WD mass qualifies them for being polluted by s-process material, but which are not. For 14~Aur~C, apart from the difficulty imprinted on the analysis  by the star rapid rotation, the cause for the absence of s-process enrichment could be its unusually short-period of 3~days, which is likely the outcome of
a common-envelope evolution \citep{Han1995}.  It is believed that this channel does not lead to substantial accretion during the common-envelope process \citep{Ricker2008}\footnote{The low-metallicity carbon dwarf HE 0024-2523, with its 3.4~d orbital period and its strong Pb overabundance, is 
a noticeable counter-example of this statement, though \protect\citep{2003AJ....125..875L}.}. Nevertheless, the recent occurence of binary interaction is revealed by the rapid rotation of the star. The same holds true for DR Dra, whose absence of s-process overabundance is a bit more puzzling, however.  The long period of DR~Dra (904~d) suggests that the rapid rotation responsible for the chromospheric activity be due to spin accretion \citep{1996MNRAS.280.1264T} rather than to tidal coupling, which requires a closer system \citep{Fekel1993}. Thus there has been some binary interaction in the recent past of DR~Dra. The absence of s-process overabundance in that star suggests that, with a mass estimated at 0.55~\Mo\ \citep{1985AJ.....90..812F}, either the WD does not originate from a TP-AGB star \citep{2000MNRAS.315..543H}, or there was no dredge-up in the TP-AGB star, or there was simply no s-process nucleosynthesis operating.

Among barium stars themselves, the level of s-process overabundances varies greatly, from the maginal level of HR~5692, $\zeta$~Cyg, and 56~Peg, 
to the largest level of IT~Vir.   Two co-existing causes for these variations are readily identified in the literature: on the one hand, the efficiency of the s-process nucleosynthesis in AGB stars varies with metallicity  \citep{1988MNRAS.234....1C,Kappeler2011} and on the other hand, the efficiency 
of mass accretion varies with orbital separation \citep{Jorissen-Boffin-92,Han1995,2003ASPC..303..290P}. Thus, orbital separation and metallicity could directly control the level of s-process overabundances in barium stars. But all previous studies have concluded that if any, the correlation between orbital periods and s-process overabundances is weak, and the present study is no exception. For instance, 56~Peg has one of the shortest orbital periods\footnote{We implicitly assume here that the range of system masses is moderate enough for orbital periods to be a good proxy for the orbital separations.} among barium stars \citep[111~d;][]{Griffin2006} and yet its s-process overabundance level is very moderate. 
A search for extra parameters (besides metallicity) blurring the relationship between orbital periods and 
s-process pollution levels is therefore of interest. Here we suggest that the WD mass, tracing the maximum luminosity reached on the AGB \citep[through the core-mass -- luminosity relationship;][]{1998A&A...340L..43H}, could also play a role in that respect. Indeed, the AGB luminosity is a proxy for the number of thermal pulses experienced by the AGB star (although the initial AGB mass must play a role as well), which in turn governs the s-process level attained in its atmosphere.
However, the number  of stars available is still too scarce to be able to draw any firm conclusion, apart from the observation that large s-process overabundance levels are found at both ends of the CO WD mass range: around 0.6~\Mo\ (IT~Vir, V1261~Ori) and around 1~\Mo\ ($\zeta$~Cap).

\section{Conclusion}
\label{Sect:conclusion}

We have analyzed a sample of 11 binary systems comprising WD of known masses. High-resolution and high-signal-to-noise spectra were collected from  the  HERMES spectrograph installed at the Mercator 1.2m telescope. The atmospheric parameters and the CNO and s-process abundances were derived using state-of-the-art LTE models and line lists. We derived for the first time atmospheric parameters for 14~Aur~C (HD~33959C) and s-process element abundances for 56~Peg (HR~8796). The sample includes 2 S stars which have been analysed with new dedicated S-star \marcs\ model atmospheres. One of them, IT~Vir (HD~121447), has the largest overabundance of s-process elements in the sample and has been identified as highly N-enriched. Such a high level of N overabundance could perhaps be related to the photometric variations exhibited by that star, and attributed to ellipsoidal variations as the red giant has a large Roche-filling factor. The ensuing faster rotation could trigger deep mixing which is known to bring nitrogen to the surface.

Finally, from our homogeneous  study of s-process abundances in the primary components of binary systems involving a WD companion,
we find that there is a clear relationship between s-process enrichment and WD mass: as expected, primary components in binary systems involving He WDs (thus with a mass lower than about 0.5~\Mo) never show s-process overabundances, as the WD progenitor never reached the TP-AGB. 

\begin{acknowledgement}
This research has been funded by the Belgian Science Policy Office under contract BR/143/A2/STARLAB.
T.M. is supported by the FNRS-F.R.S. as temporary post-doctoral researcher  under grant No. 2.4513.11. 
This work was supported by the Fonds de la Recherche Scientifique FNRS under Grant n$\circ$ T.0198.13.
The Mercator telescope is operated thanks to grant number G.0C31.13 of the FWO under the ’Big Science’ initiative of the Flemish governement. Based on observations obtained with the HERMES spectrograph, supported by the Fund for Scientific Research of Flanders (FWO), the Research Council of K.U.Leuven, the Fonds National de la Recherche Scientifique (F.R.S.-FNRS), Belgium, the Royal Observatory of Belgium, the Observatoire de Gen\`eve, Switzerland and the Th\"uringer Landessternwarte Tautenburg, Germany. This work has made use of the VALD database, operated at Uppsala University, the Institute of Astronomy RAS in Moscow, and the University of Vienna.
\end{acknowledgement}

\bibliographystyle{aa}
\bibliography{biblio}

\onecolumn
\appendix

\section{Atomic linelist}
\label{ap:ll}
\begin{longtab}
 \begin{longtable}{lcrl}
\caption{Atomic line list. A vertical mark spanning different spectral lines identify hyperfine or isotopic components. 
\label{Tab:all}}\\
\tiny
 $\lambda$ [nm] & $\chi$ [eV] & $\log{gf}$ \\
 \hline
 \endfirsthead
 \caption{Continued.}\\
 $\lambda$ [nm] & $\chi$ [eV] & $\log{gf}$ \\
 \hline
 \endhead
 \hline
 \endfoot
\\
\textbf{C I} & \multicolumn{2}{l}{A$_\odot$(C) = 8.39}\\
538.0337 &  7.685 & $-1.615$  \\
555.3174 &  8.643 & $-2.370$  \\
768.5190 &  8.771 & $-1.519$  \\
833.5148 &  7.685 & $-0.420$  \\
872.7126 &  1.264 & $-8.136$  \\
\\
\textbf{O I} & \multicolumn{2}{l}{A$_\odot$(O) = 8.66}\\
 615.6776 & 10.741 & $ -0.694$ \\
 630.0304 &  0.000 & $ -9.715$ \\
 636.3776 &  0.020 & $-10.190$ \\
 777.1941 &  9.146 & $  0.369$ &only in IT~Vir \\ 
 777.4161 &  9.146 & $  0.223$ &only in IT~Vir \\
 777.5388 &  9.146 & $  0.001$ &only in IT~Vir \\
 844.6359 &  9.521 & $  0.236$ \\
 844.6758 &  9.521 & $  0.014$ \\
\\
\textbf{Sr I} & \multicolumn{2}{l}{A$_\odot$(Sr) = 2.92}\\
460.7327 &  0.000 & $ 0.283$ & NIST AA \\
640.8459 &  2.271 & $ 0.510$ & in the red wing of a \ion{Fe}{i} line \\
687.8310 &  1.798 & $-0.240$ & only in V1261~Ori\\
689.2581 &  0.000 & $-2.840$ & only in V1261~Ori\\
707.0070 &  1.847 & $-0.030$ &                  \\
\\
\textbf{Sr II} & \multicolumn{2}{l}{A$_\odot$(Sr) = 2.92}\\
407.7719 & 0.000 & $ 0.170$ & only in 14 Aur C \\
416.1792 & 2.940 & $-0.600$ & only in 14 Aur C and 63~Eri \\
421.5519 & 0.000 & $-0.170$ & only in 14 Aur C \\
\\
\textbf{Y I} & \multicolumn{2}{l}{A$_\odot$(Y) = 2.21}\\
563.0130 &  1.356 & $ 0.211$ & only in IT Vir\\
613.8435 &  0.066 & $-1.923$ & \\
622.2578 &  0.000 & $-1.452$ & \\
640.2006 &  0.066 & $-1.849$ & only in 56~Peg and IT Vir\\
643.5004 &  0.066 & $-0.820$ & \\
668.7567 &  0.000 & $-2.000$ & only in V1261~Ori \\
\\
\textbf{Y II} & \multicolumn{2}{l}{A$_\odot$(Y) = 2.21}\\
395.0352 &  0.104 &  $-0.490$ & only in 14~Aur~C\\
488.3684 &  1.084 &  $ 0.265$ & only in 14~Aur~C and V1379 Aql\\
490.0120 &  1.033 &  $ 0.103$ & only in IP~Eri\\
508.7416 &  1.084 &  $-0.170$ & \\
520.0406 &  0.992 &  $-0.570$ & \\
528.9815 &  1.033 &  $-1.850$ & \\
532.0782 &  1.084 &  $-1.950$ & only in V1379 Aql\\
540.2774 &  1.839 &  $-0.630$ & \\
554.4611 &  1.738 &  $-1.090$ & only in 63~Eri and V1379 Aql\\
554.6009 &  1.748 &  $-0.754$ & \\
566.2925 &  1.944 &  $ 0.384$ & \\
572.8890 &  1.839 &  $-1.120$ & \\ 
661.3733 &  1.748 &  $-0.848$ & \\
679.5414 &  1.738 &  $-1.030$ & \\
788.1881 &  1.839 &  $-0.570$ & \\
\\
\textbf{Zr I} & \multicolumn{2}{l}{A$_\odot$(Zr) = 2.58}\\
473.9480 & 0.651 & $ 0.230$ & only in 63~Eri \\ 
477.2323 & 0.623 & $ 0.040$ & \\
488.7750 & 0.730 & $ 1.000$ & \\
538.5151 & 0.519 & $-0.710$ & \\
568.0.92 & 0.543 & $-1.700$ & only in IT Vir\\
573.5690 & 0.000 & $-2.240$ & only in $\zeta$~Cap\\
612.7475 & 0.154 & $-1.060$ & \\
613.4585 & 0.000 & $-1.280$ & \\
614.0535 & 0.519 & $-1.410$ & \\
614.3252 & 0.071 & $-1.100$ & \\
644.5747 & 0.999 & $-0.830$ & \\
650.6437 & 0.633 & $-2.110$ & only in $\zeta$~Cap\\
699.0869 & 0.623 & $-1.220$ & \\
709.7774 & 0.687 & $-0.570$ & only in V1379 Aql \\
716.9130 & 0.730 & $-0.880$ & \\
733.6066 & 0.519 & $-2.170$ & only in IT Vir \\ 
743.9889 & 0.543 & $-1.810$ & \\
755.3039 & 0.519 & $-3.040$ & only in V1261~Ori\\
755.4780 & 0.520 & $-2.280$ & only in V1261~Ori\\
755.8487 & 1.550 & $-1.470$ & only in V1261~Ori\\
760.7167 & 0.633 & $-1.880$ & \\
784.9365 & 0.687 & $-1.300$ & \\
805.8107 & 0.623 & $-2.020$ & only in IT Vir\\
806.3105 & 0.623 & $-1.620$ & only in $\zeta$~Cap\\
807.0115 & 0.730 & $-0.790$ & only in IP~Eri and IT Vir\\
813.3011 & 0.687 & $-1.130$ & only in $\zeta$~Cap and IT Vir\\
846.4687 & 0.651 & $-2.080$ & only in IT Vir \\
\\
\textbf{Zr II} & \multicolumn{2}{l}{A$_\odot$(Zr) = 2.58}\\
4150.986 & 0.802 & $-0.992$ & only in 14~Aur~C\\
4211.877 & 0.527 & $-1.040$ & only in 14~Aur~C\\
4379.742 & 1.532 & $-0.356$ & only in 14~Aur~C\\
5112.270 & 1.665 & $-0.850$ & \\
5350.350 & 1.773 & $-1.160$ & \\
5418.016 & 1.756 & $-1.600$ & only in HR5692\\
\\
\textbf{Ba I} & \multicolumn{2}{l}{A$_\odot$(Ba) = 2.17}\\
 748.8077 & 1.190 & $-0.230$ & \\ 
\\ 
\textbf{Ba II} & \multicolumn{2}{l}{A$_\odot$(Ba) = 2.17}\\
 \vline~389.1776$^{134}$ & 2.512 & $ 0.295$ & \\
 \vline~389.1776$^{135}$ & 2.512 & $ 0.295$ & \\
 \vline~389.1776$^{136}$ & 2.512 & $ 0.295$ & Only in HR5692 \\
 \vline~389.1776$^{137}$ & 2.512 & $ 0.295$ & \\
 \vline~389.1776$^{138}$ & 2.512 & $ 0.295$ & \\
 \\
 \vline~416.6000$^{134}$ & 2.722 & $-0.433$ & \\
 \vline~416.6000$^{135}$ & 2.722 & $-0.433$ & \\
 \vline~416.6000$^{136}$ & 2.722 & $-0.433$ & Only in DR Dra\\
 \vline~416.6000$^{137}$ & 2.722 & $-0.433$ & \\
 \vline~416.6000$^{138}$ & 2.722 & $-0.433$ & \\ 
 \\
 \vline~452.4925$^{130}$ & 2.512 & $-0.390$ & \\
 \vline~452.4925$^{132}$ & 2.512 & $-0.390$ & \\
 \vline~452.4925$^{134}$ & 2.512 & $-0.390$ & \\
 \vline~452.4925$^{135}$ & 2.512 & $-0.390$ & Only in 63 Eri\\
 \vline~452.4925$^{136}$ & 2.512 & $-0.390$ & \\
 \vline~452.4925$^{137}$ & 2.512 & $-0.390$ & \\
 \vline~452.4925$^{138}$ & 2.512 & $-0.390$ & \\

 \\
\vline~455.3998$^{137}$ & 0.000 & $-0.666$ & \\
\vline~455.3999$^{137}$ & 0.000 & $-0.666$ & \\
\vline~455.4000$^{137}$ & 0.000 & $-1.064$ & \\
\vline~455.4001$^{135}$ & 0.000 & $-0.666$ & \\
\vline~455.4002$^{135}$ & 0.000 & $-1.064$ & \\
\vline~455.4002$^{135}$ & 0.000 & $-0.666$ & \\
\vline~455.4031$^{130}$ & 0.000 & $ 0.140$ & \\
\vline~455.4031$^{132}$ & 0.000 & $ 0.140$ & \\
\vline~455.4031$^{134}$ & 0.000 & $ 0.140$ & Only in 14 Aur C and V1379 Aql\\
\vline~455.4032$^{136}$ & 0.000 & $ 0.140$ & \\
\vline~455.4033$^{138}$ & 0.000 & $ 0.140$ & \\
\vline~455.4048$^{135}$ & 0.000 & $-0.219$ & \\
\vline~455.4050$^{135}$ & 0.000 & $-0.666$ & \\
\vline~455.4051$^{137}$ & 0.000 & $-0.219$ & \\
\vline~455.4052$^{135}$ & 0.000 & $-1.365$ & \\
\vline~455.4054$^{137}$ & 0.000 & $-0.666$ & \\
\vline~455.4055$^{137}$ & 0.000 & $-1.365$ & \\
\\
\vline~493.4030$^{137}$ & 0.000 & $-0.662$ & \\ 
\vline~493.4032$^{135}$ & 0.000 & $-0.662$ & \\ 
\vline~493.4042$^{135}$ & 0.000 & $-1.361$ & \\ 
\vline~493.4042$^{137}$ & 0.000 & $-1.361$ & \\ 
\vline~493.4074$^{130}$ & 0.000 & $-0.157$ & \\ 
\vline~493.4074$^{132}$ & 0.000 & $-0.157$ & \\ 
\vline~493.4074$^{134}$ & 0.000 & $-0.157$ & Only in 14 Aur C \\ 
\vline~493.4075$^{136}$ & 0.000 & $-0.157$ & \\ 
\vline~493.4077$^{138}$ & 0.000 & $-0.157$ & \\ 
\vline~493.4091$^{135}$ & 0.000 & $-0.662$ & \\ 
\vline~493.4095$^{137}$ & 0.000 & $-0.662$ & \\ 
\vline~493.4102$^{135}$ & 0.000 & $-0.662$ & \\ 
\vline~493.4107$^{137}$ & 0.000 & $-0.662$ & \\ 
\\
\vline~585.3669$^{135}$ & 0.604 & $-1.967$ & \\
\vline~585.3669$^{137}$ & 0.604 & $-1.967$ & \\
\vline~585.3670$^{135}$ & 0.604 & $-2.113$ & \\
\vline~585.3670$^{135}$ & 0.604 & $-1.909$ & \\
\vline~585.3671$^{137}$ & 0.604 & $-2.113$ & \\
\vline~585.3671$^{137}$ & 0.604 & $-1.909$ & \\
\vline~585.3672$^{135}$ & 0.604 & $-2.113$ & \\
\vline~585.3672$^{135}$ & 0.604 & $-2.511$ & \\
\vline~585.3673$^{130}$ & 0.604 & $-0.909$ & \\
\vline~585.3673$^{132}$ & 0.604 & $-0.909$ & \\
\vline~585.3673$^{134}$ & 0.604 & $-0.909$ & \\
\vline~585.3673$^{137}$ & 0.604 & $-2.113$ & \\
\vline~585.3673$^{135}$ & 0.604 & $-1.812$ & \\
\vline~585.3673$^{137}$ & 0.604 & $-2.511$ & \\
\vline~585.3674$^{136}$ & 0.604 & $-0.909$ & \\
\vline~585.3675$^{135}$ & 0.604 & $-1.909$ & \\
\vline~585.3675$^{135}$ & 0.604 & $-1.365$ & \\
\vline~585.3675$^{137}$ & 0.604 & $-1.812$ & \\
\vline~585.3675$^{138}$ & 0.604 & $-0.909$ & \\
\vline~585.3676$^{137}$ & 0.604 & $-1.909$ & \\
\vline~585.3676$^{137}$ & 0.604 & $-1.365$ & \\
\vline~585.3680$^{135}$ & 0.604 & $-1.967$ & \\
\vline~585.3682$^{137}$ & 0.604 & $-1.967$ & \\
\\
\vline~614.1708$^{135}$ & 0.704 & $-0.456$ & \\
\vline~614.1708$^{135}$ & 0.704 & $-1.264$ & \\
\vline~614.1709$^{135}$ & 0.704 & $-2.410$ & \\
\vline~614.1709$^{137}$ & 0.704 & $-1.264$ & \\
\vline~614.1709$^{137}$ & 0.704 & $-0.456$ & \\
\vline~614.1710$^{137}$ & 0.704 & $-2.410$ & \\
\vline~614.1711$^{130}$ & 0.704 & $-0.030$ & \\
\vline~614.1711$^{132}$ & 0.704 & $-0.030$ & \\
\vline~614.1711$^{134}$ & 0.704 & $-0.030$ & \\
\vline~614.1712$^{136}$ & 0.704 & $-0.030$ & \\
\vline~614.1713$^{135}$ & 0.704 & $-0.662$ & \\
\vline~614.1713$^{138}$ & 0.704 & $-0.030$ & \\
\vline~614.1714$^{135}$ & 0.704 & $-1.167$ & \\
\vline~614.1715$^{135}$ & 0.704 & $-2.234$ & \\
\vline~614.1715$^{137}$ & 0.704 & $-0.662$ & \\
\vline~614.1716$^{135}$ & 0.704 & $-0.912$ & \\
\vline~614.1716$^{137}$ & 0.704 & $-1.167$ & \\
\vline~614.1717$^{135}$ & 0.704 & $-1.234$ & \\
\vline~614.1717$^{135}$ & 0.704 & $-1.280$ & \\
\vline~614.1717$^{137}$ & 0.704 & $-2.234$ & \\
\vline~614.1718$^{137}$ & 0.704 & $-0.912$ & \\
\vline~614.1719$^{137}$ & 0.704 & $-1.234$ & \\
\vline~614.1719$^{137}$ & 0.704 & $-1.280$ & \\
\\
\vline~649.6883$^{135}$ & 0.604 & $-1.911$ & \\
\vline~649.6883$^{137}$ & 0.604 & $-1.911$ & \\
\vline~649.6888$^{135}$ & 0.604 & $-1.212$ & \\
\vline~649.6888$^{137}$ & 0.604 & $-1.212$ & \\
\vline~649.6895$^{130}$ & 0.604 & $-0.406$ & \\
\vline~649.6895$^{132}$ & 0.604 & $-0.406$ & \\
\vline~649.6895$^{134}$ & 0.604 & $-0.406$ & \\
\vline~649.6895$^{135}$ & 0.604 & $-0.765$ & \\
\vline~649.6896$^{137}$ & 0.604 & $-0.765$ & \\
\vline~649.6897$^{136}$ & 0.604 & $-0.406$ & \\
\vline~649.6898$^{138}$ & 0.604 & $-0.406$ & \\
\vline~649.6900$^{135}$ & 0.604 & $-1.610$ & \\
\vline~649.6902$^{135}$ & 0.604 & $-1.212$ & \\
\vline~649.6902$^{137}$ & 0.604 & $-1.610$ & \\
\vline~649.6904$^{137}$ & 0.604 & $-1.212$ & \\
\vline~649.6906$^{135}$ & 0.604 & $-1.212$ & \\
\vline~649.6909$^{137}$ & 0.604 & $-1.212$ & \\
\\
\textbf{La II} & \multicolumn{2}{l}{A$_\odot$(La) = 1.13}\\
\vline~394.9037 & 0.403 & $-1.337$ & \\
\vline~394.9038 & 0.403 & $-1.191$ & \\
\vline~394.9044 & 0.403 & $-0.995$ & \\
\vline~394.9046 & 0.403 & $-1.008$ & \\
\vline~394.9047 & 0.403 & $-1.668$ & \\
\vline~394.9056 & 0.403 & $-0.761$ & \\
\vline~394.9058 & 0.403 & $-0.886$ & \\
\vline~394.9059 & 0.403 & $-1.559$ & \\
\vline~394.9072 & 0.403 & $-0.576$ & \\
\vline~394.9075 & 0.403 & $-0.825$ &  Only in 14 Aur C\\
\vline~394.9077 & 0.403 & $-1.580$ & \\
\vline~394.9093 & 0.403 & $-0.420$ & \\
\vline~394.9097 & 0.403 & $-0.821$ & \\
\vline~394.9100 & 0.403 & $-1.690$ & \\
\vline~394.9120 & 0.403 & $-0.284$ & \\
\vline~394.9124 & 0.403 & $-0.887$ & \\
\vline~394.9127 & 0.403 & $-1.902$ & \\
\vline~394.9151 & 0.403 & $-0.163$ & \\
\vline~394.9156 & 0.403 & $-1.092$ & \\
\vline~394.9160 & 0.403 & $-2.305$ & \\
\\
\vline~398.8445 & 0.403 & $-1.362$ & \\
\vline~398.8446 & 0.403 & $-1.839$ & \\
\vline~398.8452 & 0.403 & $-1.140$ & \\
\vline~398.8454 & 0.403 & $-1.362$ & \\
\vline~398.8464 & 0.403 & $-1.041$ & \\
\vline~398.8466 & 0.403 & $-1.985$ & \\
\vline~398.8467 & 0.403 & $-1.140$ & \\
\vline~398.8482 & 0.403 & $-1.015$ & \\
\vline~398.8484 & 0.403 & $-1.355$ & \\
\vline~398.8486 & 0.403 & $-1.041$ & Only in 63 Eri\\
\vline~398.8504 & 0.403 & $-1.068$ & \\
\vline~398.8507 & 0.403 & $-0.969$ & \\
\vline~398.8509 & 0.403 & $-1.015$ & \\
\vline~398.8532 & 0.403 & $-1.263$ & \\
\vline~398.8535 & 0.403 & $-0.683$ & \\
\vline~398.8538 & 0.403 & $-1.068$ & \\
\vline~398.8569 & 0.403 & $-0.455$ & \\
\vline~398.8572 & 0.403 & $-1.263$ & \\
\\
404.2901 & 0.927 & 0.290 & Only in 63 Eri \\
\\
\vline~408.6695 & 0.000 & $-1.266$ & \\
\vline~408.6699 & 0.000 & $-1.108$ & \\
\vline~408.6702 & 0.000 & $-1.119$ & \\
\vline~408.6705 & 0.000 & $-1.292$ & \\
\vline~408.6708 & 0.000 & $-0.696$ & \\
\vline~408.6709 & 0.000 & $-1.094$ & \\
\vline~408.6710 & 0.000 & $-1.790$ & Only in HR 5692\\
\vline~408.6711 & 0.000 & $-1.468$ & \\
\vline~408.6711 & 0.000 & $-3.216$ & \\
\vline~408.6717 & 0.000 & $-1.292$ & \\
\vline~408.6719 & 0.000 & $-1.119$ & \\
\vline~408.6720 & 0.000 & $-1.108$ & \\
\vline~408.6721 & 0.000 & $-1.266$ & \\
\\
\vline~466.2478 & 0.000 & $-2.952$ & \\
\vline~466.2482 & 0.000 & $-2.511$ & \\
\vline~466.2486 & 0.000 & $-2.240$ & \\
\vline~466.2491 & 0.000 & $-2.253$ & \\
\vline~466.2492 & 0.000 & $-2.137$ & \\
\vline~466.2493 & 0.000 & $-2.256$ & \\
\vline~466.2503 & 0.000 & $-2.511$ & \\
\vline~466.2505 & 0.000 & $-2.056$ & \\
\vline~466.2507 & 0.000 & $-1.763$ & \\
\\
474.8726 & 0.927 & $-0.540$ & Only in $\zeta$ Cyg and 63 Eri \\
480.4039 & 0.235 & $-1.490$ & Only in 56 Peg and IT vir\\
\\
\vline~492.1774 & 0.244 & $-1.139$ & \\
\vline~492.1774 & 0.244 & $-2.220$ & \\
\vline~492.1774 & 0.244 & $-3.601$ & \\
\vline~492.1775 & 0.244 & $-1.233$ & \\
\vline~492.1775 & 0.244 & $-2.005$ & \\
\vline~492.1775 & 0.244 & $-3.207$ & \\
\vline~492.1776 & 0.244 & $-1.334$ & \\
\vline~492.1776 & 0.244 & $-1.445$ & \\
\vline~492.1776 & 0.244 & $-1.915$ & \\
\vline~492.1776 & 0.244 & $-1.927$ & \\
\vline~492.1776 & 0.244 & $-2.923$ & Only in $\zeta$ Cap et 14 Aur C\\
\vline~492.1776 & 0.244 & $-3.010$ & \\
\vline~492.1777 & 0.244 & $-1.566$ & \\
\vline~492.1777 & 0.244 & $-1.955$ & \\
\vline~492.1777 & 0.244 & $-2.939$ & \\
\vline~492.1778 & 0.244 & $-1.700$ & \\
\vline~492.1778 & 0.244 & $-1.848$ & \\
\vline~492.1778 & 0.244 & $-2.006$ & \\
\vline~492.1778 & 0.244 & $-2.053$ & \\
\vline~492.1778 & 0.244 & $-2.258$ & \\
\vline~492.1778 & 0.244 & $-3.123$ & \\
\\
511.4559 & 0.235 & $-1.030$ & Only in 14 Aur C\\
529.0818 & 0.000 & $-1.650$ & \\
\\
\vline~530.1908 & 0.403 & $-3.065$ & \\
\vline~530.1913 & 0.403 & $-2.266$ & \\
\vline~530.1917 & 0.403 & $-2.391$ & \\
\vline~530.1946 & 0.403 & $-3.483$ & \\
\vline~530.1953 & 0.403 & $-2.300$ & \\
\vline~530.1958 & 0.403 & $-2.120$ & \\
\vline~530.2001 & 0.403 & $-2.483$ & \\
\vline~530.2008 & 0.403 & $-1.913$ & \\
\vline~530.2067 & 0.403 & $-1.742$ & \\
\\
\vline~530.3513 & 0.321 & $-1.874$ & \\ 
\vline~530.3513 & 0.321 & $-2.363$ & \\ 
\vline~530.3514 & 0.321 & $-3.062$ & \\ 
\vline~530.3531 & 0.321 & $-2.167$ & \\ 
\vline~530.3532 & 0.321 & $-2.247$ & \\ 
\vline~530.3532 & 0.321 & $-2.622$ & \\ 
\vline~530.3546 & 0.321 & $-2.366$ & \\ 
\vline~530.3546 & 0.321 & $-2.622$ & \\ 
\vline~530.3547 & 0.321 & $-2.351$ & \\ 
\\
580.5773 & 0.126 & $-1.560$ & \\
586.3691 & 0.927 & $-1.370$ & Only in $\zeta$ Cap\\
593.6210 & 0.173 & $-2.070$ & \\
617.2721 & 0.126 & $-2.253$ & \\
\\
\vline~626.2422 & 0.403 & $-1.873$ & \\
\vline~626.2429 & 0.403 & $-2.802$ & \\
\vline~626.2434 & 0.403 & $-4.015$ & \\
\\
\vline~639.0455 & 0.321 & $-2.012$ & \\
\vline~639.0468 & 0.321 & $-2.183$ & \\
\vline~639.0468 & 0.321 & $-2.752$ & \\
\vline~639.0479 & 0.321 & $-2.570$ & \\
\vline~639.0479 & 0.321 & $-3.752$ & \\
\vline~639.0480 & 0.321 & $-2.390$ & \\
\vline~639.0489 & 0.321 & $-2.536$ & \\
\vline~639.0489 & 0.321 & $-3.334$ & \\
\vline~639.0490 & 0.321 & $-2.661$ & \\
\vline~639.0496 & 0.321 & $-3.100$ & \\
\vline~639.0497 & 0.321 & $-2.595$ & \\
\vline~639.0498 & 0.321 & $-3.079$ & \\
\vline~639.0502 & 0.321 & $-2.954$ & \\
\vline~639.0503 & 0.321 & $-2.778$ & \\
\vline~639.0506 & 0.321 & $-2.857$ & \\
\\
677.4268 & 0.126 & $-1.708$ & \\
683.4099 & 0.244 & $-2.168$ & Only in $\zeta$ Cap and 56 Peg\\
706.6198 & 0.000 & $-1.655$ & Only in IP Eri and DR Dra\\
748.3490 & 0.126 & $-1.891$ & \\
\\
\textbf{Ce II} & \multicolumn{2}{l}{A$_\odot$(Ce) = 1.70}\\
 \vline~434.9768 & 0.529 & $-0.520$ & \\
 \vline~434.9789 & 0.701 & $-0.350$ & \\
 451.5848 & 1.058 & $-0.240$ & Only in $\zeta$ Cyg and HR 5692\\
 456.2359 & 0.478 & $ 0.230$ & \\
 462.8239 & 1.366 & $-0.430$ & \\
 477.3941 & 0.924 & $-0.390$ & Only in $\zeta$ Cap \\
 494.3441 & 1.206 & $-0.360$ & \\
 494.4618 & 1.008 & $-0.520$ & Only in $\zeta$ Cap \\
 518.7503 & 0.559 & $-3.330$ & \\
 527.4229 & 1.044 & $ 0.130$ & Only in $\zeta$ Cap and 14 Aur C\\
 533.0556 & 0.869 & $-0.400$ & \\
 5472.279 & 1.247 & $-0.100$ & Only in 63 Eri\\
 \vline~597.5818 & 1.327 & $-0.460$ & Only in IP Eri\\
 \vline~597.5881 & 1.251 & $-1.680$ & and HR 5692\\
 604.3373 & 1.206 & $-0.480$ & \\
 605.1815 & 0.232 & $-1.530$ & \\
 605.2617 & 1.838 & $-0.800$ & Only in $\zeta$ Cap\\
 758.0913 & 0.327 & $-2.120$ & Only in V1261 Ori \\
 802.5571 & 0.000 & $-1.420$ & \\
 871.6659 & 0.122 & $-1.980$ & \\
 877.2135 & 0.357 & $-1.260$ & \\
 891.0948 & 0.435 & $-1.800$ & \\
 897.0176 & 0.295 & $-2.020$ & Only in V1261 Ori \\
 899.3403 & 0.265 & $-2.520$ & Only in V1261 Ori \\
 \\
 \hline
\end{longtable}
\normalsize
\end{longtab}

\section{Molecular linelist}
\label{ap:ml}
\begin{longtab}
 \begin{longtable}{ll}
\caption{\label{Tab:mll}
Molecular linelist.}\\
\tiny
 $\lambda$ [nm] & comments \\
 \hline
 \endfirsthead
 \caption{Continued.}\\
 $\lambda$ [nm]  \\
 \hline
 \endhead
 \hline
 \endfoot
\\
\textbf{C$_2$} & \\
468.427 & only used for 14 Aur C \\
471.30  & \\
473.277 & \\
473.50  & \\
473.531 & \\
489.748 & region difficult to fit\\
493.667 & \\
495.142 & \\
495.60  & \\
496.363 & \\
502.416 & \\
503.36  & \\
505.88  & \\
506.672 & \\
507.35  & \\
507.528 & \\
508.92  & \\
513.56  & used in \citet{Barbuy1992}\\
514.12  & \\
558.52  & \\
562.12  & \\
\\
\textbf{CN} & \\
639.63 & \\
642.07 & \\
645.85 & \\
647.72 & \\
651.97 & \\
653.56 & \\
654.90 & \\
696.19 & \\
697.42 & \\
700.91 & \\
700.96 & \\
701.19 & \\
720.52 & \\
723.81 & \\
726.36 & \\
733.70 & \\
734.01 & \\
734.06 & \\
734.23 & \\
737.45 & \\
738.45 & \\
739.79 & \\
740.34 & \\
745.63 & \\
749.92 & \\
753.95 & \\
754.91 & \\
757.19 & \\
787.40 & \\
789.21 & \\
790.67 & \\
790.96 & \\
791.18 & \\
792.30 & \\
792.97 & \\
793.86 & \\
793.92 & \\
795.17 & \\
795.41 & \\
795.82 & \\
796.71 & \\
797.08 & \\
797.38 & \\
797.47 & \\
797.72 & \\
797.98 & \\
798.11 & \\
798.46 & \\
798.52 & \\
799.23 & \\
799.56 & \\
800.85 & \\
801.57 & \\
801.70 & \\
801.81 & \\
803.28 & \\
803.50 & \\
803.81 & \\
804.17 & \\
805.31 & \\
805.73 & \\
806.41 & \\
807.44 & \\
808.25 & \\
808.62 & \\
808.73 & \\
808.83 & \\
810.04 & \\
810.15 & \\
811.87 & \\
812.40 & \\
812.56 & \\
813.21 & \\
816.01 & \\
816.69 & \\
817.51 & \\
818.02 & \\
820.70 & \\
823.60 & \\
824.73 & \\
826.72 & \\
826.80 & \\
829.12 & \\
831.24 & \\
832.55 & \\
833.04 & \\
833.74 & \\
835.07 & \\
837.15 & \\
839.12 & \\
839.85 & \\
840.72 & \\
841.13 & \\
842.05 & \\
842.24 & \\
843.00 & \\
843.12 & \\
844.18 & \\
845.78 & \\
846.99 & \\
847.09 & \\
847.24 & \\
847.84 & \\
850.51 & \\
852.34 & \\
856.70 & \\
857.33 & \\
857.72 & \\
858.70 & \\
860.49 & \\
860.77 & \\
862.05 & \\
862.44 & \\
863.00 & \\
862.91 & \\
870.31 & \\
870.68 & \\
875.97 & \\
881.62 & \\
881.79 & \\
884.19 & \\
\hline
\end{longtable}
\normalsize
\end{longtab}

\end{document}